\documentclass[useAMS]{mn2e}
\usepackage{epsfig}
\usepackage{lscape}
\newcommand{\beq}{\begin{equation}}
\newcommand{\eeq}{\end{equation}}

\def\lsim{\mathrel{\lower2.5pt\vbox{\lineskip=0pt\baselineskip=0pt
           \hbox{$<$}\hbox{$\sim$}}}}
\def\gsim{\mathrel{\lower2.5pt\vbox{\lineskip=0pt\baselineskip=0pt
           \hbox{$>$}\hbox{$\sim$}}}}
\def\Lsun{\hbox{L$_{\odot}$}}
\def\Msun{\hbox{M$_{\odot}$}}

\input epsf

\begin{document}


\title[AzTEC millimeter survey of the COSMOS field - III.]
{AzTEC millimeter survey of the COSMOS field - III. Source catalog over
0.72 sq. deg. and plausible boosting by large-scale structure}
\author[Aretxaga I. et al.]
{I. Aretxaga$^1$, G.W. Wilson$^2$, E. Aguilar$^1$,  S. Alberts$^2$,
K. S. Scott$^3$, N. Scoville$^4$,
\newauthor
 M.S. Yun$^{2}$,  J. Austermann$^{5}$, T.P. Downes$^{4}$,
H. Ezawa$^{6}$, B. Hatsukade$^{7}$,
\newauthor
D.H. Hughes$^{1}$, R. Kawabe$^{7}$, K. Kohno$^{8}$, T. Oshima$^7$, T.A. Perera$^{9}$,
Y. Tamura$^7$,
\newauthor
M. Zeballos$^{1}$
\\
$^1$Instituto Nacional de Astrof\'{\i}sica, \'Optica y Electr\'onica
(INAOE), Aptdo. Postal 51 y 216, 72000 Puebla, Pue., Mexico  \\
$^2$Department of Astronomy, University of Massachusetts, Amherst, MA 01003, USA.\\
$^3$  Department of Physics and Astronomy, University of Pennsylvania, Philadephia, PA 19104,
USA\\
$^4$  California Institute of Technology, Pasadena, CA 91125, USA\\
$^5$ Center for Astrophysics and Space Astronomy, University of Colorado, Boulder, CO 80309, USA\\
$^6$ ALMA Project Office, National Astronomical Observatory of Japan,
2-21-1 Osawa, Mitaka, Tokyo 181-8588, Japan\\
$^7$  Nobeyama Radio Observatory, Minamimaki, Minamisaku, Nagano 384-1305, Japan\\
$^8$ Institute of Astronomy, University of Tokyo, 2-21-1 Osawa, Mitaka, Tokyo 181-0015, Japan\\
$^9$   Department of Physics, Illinois Wesleyan University, Bloomington, IL 61701, USA\\
}


\date{}

\pagerange{} \pubyear{}

\maketitle

\label{firstpage}

\begin{abstract}
We present a $0.72$~sq. deg. contiguous 1.1mm survey in the central area
of the COSMOS field carried out to a $1\sigma \approx 1.26$~mJy
beam$^{-1}$ depth with the AzTEC camera mounted on the 10m Atacama
Submillimeter Telescope Experiment (ASTE).  We have uncovered 189 candidate
sources at a signal-to-noise ratio S/N$\geq 3.5$, out of which 129, with S/N$\geq 4$, can be
considered to have little chance of being spurious ($\lsim 2$~per cent).
We present the number counts derived with this survey, which show a
significant excess of sources when compared to the number counts derived
from the $\sim$0.5 sq. deg. area sampled at similar depths in the
Scuba HAlf Degree Extragalactic Survey (SHADES, Austermann et al. 2010).
They are, however, consistent with those derived from fields that were
considered too small to characterize the overall blank-field population.
We identify differences to
be more significant in the $S_{1.1 \rm mm}
\gsim 5$~mJy regime, and demonstrate that these excesses in number counts are
related to the areas where galaxies at redshifts $z\lsim 1.1$ are more
densely clustered. The positions of optical-IR galaxies in the redshift
interval $0.6 \lsim z \lsim 0.75$ are the ones that show the strongest
correlation
with the positions of the 1.1mm
bright population ($S_{1.1 \rm mm}\gsim 5$~mJy), a result which
does not depend
exclusively on the presence of rich clusters within the survey sampled area.
The most likely
explanation for the observed excess in number counts at 1.1mm is
galaxy-galaxy and galaxy-group lensing at moderate amplification levels,
that increases
in amplitude as one samples larger and larger flux densities.
This effect should also be detectable in other high redshift populations.
\end{abstract}


\section{Introduction}

The Cosmological Evolution Survey (COSMOS) 2~sq. deg. field has been
extensively targeted by a wide array of observations in order to probe
the cosmic evolution of galaxies and the large-scale
structure in which they are immersed (Scoville et al. 2007a). With a
wealth of multi-wavelength data spanning from X-rays to
radio-wavelengths, and a core deep UV-optical-IR survey with the
highest resolution and sensitivity offered by space facilities ({\it HST},
{\it Spitzer}, {\it GALEX}), it provides a unique opportunity to study the
relationships and interactions among galaxy populations selected at
different wavelengths and across a wide array of environments in
cosmic time.

A key contribution towards this knowledge comes from the far-IR to millimeter
wavelength regime, which has been shown to uncover ultraluminous violently
star-forming galaxies at high redshifts ($z\gsim2$) that would have
gone undetected at traditional optical-near IR survey wavelengths due
to their
intrinsic high obscuration (Smail, Ivison \& Blain 1997,
Barger et al. 1998,
Hughes et al. 1998). Named
the (sub-)millimeter galaxy population (SMG for sort), this population
has been linked to the formation of
massive elliptical galaxies, with large luminosities
$L\gsim 10^{13}$\Lsun,
large star formation rates $\gsim 1000$\Msun~yr$^{-1}$,
large reservoirs of gas $\gsim 10^{10}$\Msun, and large
dynamical $\gsim 10^{11}$\Msun\ and stellar masses $\gsim 10^{11}$\Msun\
(e.g. Greve et al. 2005, Tacconi et al. 2008,
Dye et al. 2008).

Previous mm-wavelength surveys in COSMOS have covered areas which were
significantly smaller than the full 2~sq. deg. design survey and the
conclusions derived from the limited number of detected
galaxies suffered from large field-to-field variations.  COSBO
(Bertoldi et al. 2007), a 1.2mm survey carried out with the Max-Planck
Millimeter Bolometer Array (MAMBO) mounted on the 30m Institute for
Radioastronomy at Millimeter Wavelengths (IRAM) telescope, detected 37
candidate sources at a S/N$\geq 3.5$ in the central 0.09~sq. deg.  of
COSMOS, which was mapped at a $1\sigma$ noise level of $\sim
1$~mJy~beam$^{-1}$. An adjacent 0.15~sq. deg. field was imaged at
1.1mm to a 1$\sigma$ level of 1.2 to 1.4~mJy~beam$^{-1}$ with the
AzTEC instrument (Wilson et al. 2008a)
mounted on the 15m James Clerk Maxwell Telescope (JCMT), uncovering 50
candidate sources at a S/N$\geq 3.5$ (Scott et al. 2008).
The combined area sampled by these two surveys amounts to only $\sim
12$~per cent of the total $1^{\rm o}.4\times1^{\rm o}.4$ COSMOS
area.

Field-to-field variations in the derived
overall properties of SMGs, such as number
counts, redshift distributions, or clustering, had already been reported
in these kinds of smaller ($\lsim0.25$ sq. deg.) fields and attributed
both to variance due to the intrinsically small volume sampled by the
surveys (e.g. Coppin et al. 2006,
Aretxaga et al. 2007,  Wei$\beta$ et al. 2009, Austermann et
al. 2010) and  to the chance amplification by a foreground population of
galaxies (Almaini et al. 2005, Austermann et al. 2009). Indeed, the
environments traced by optical-IR galaxies in the COSMOS sub-fields
were quite different: while the AzTEC survey focused on an area with an
overdensity of optical-IR galaxies, COSBO sampled lower galaxy-density
environments (see
figure 1).

\begin{figure}
\epsfig{file=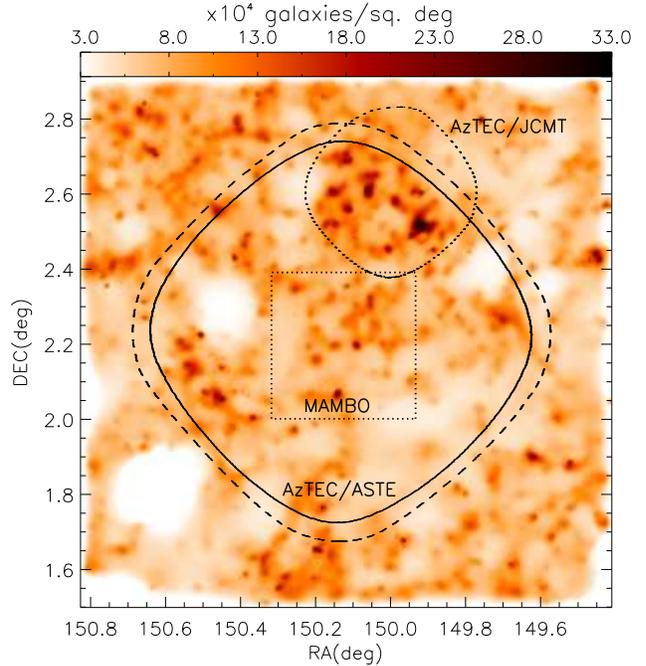,width=1.1\hsize,angle=90}
\caption{Representation
of the optically-IR selected galaxy density map
in the COSMOS field
(Scoville et al. 2007b) and the uniform coverage areas of the
mm surveys published to date: the MAMBO COSBO survey at 1.2mm to
a $1\sigma$ noise level of
$\sim 1$~mJy~beam$^{-1}$ (Bertoldi et al. 2007);
and the AzTEC/JCMT survey at 1.1mm to a 1$\sigma$ level of 1.2 to
1.4~mJy~beam$^{-1}$ (Scott et al. 2008). The
AzTEC/ASTE survey at 1.1mm, presented in this paper, has an
average $1\sigma$ noise level of 1.26~mJy~beam$^{-1}$ within the
0.72~sq. deg. uniform coverage area (solid line). Also represented is
a concentric area that marks the 25~per cent coverage area of the
survey, whose noise level increases towards the edges of the map.}
\label{fig:numcounts}
\end{figure}

A more representative survey of the COSMOS field was thus necessary in
order to investigate the culprits for these large variations and to
characterize the global blank-field population  at 1.1mm. AzTEC alone has
surveyed to date $\sim 2$~sq. deg. of the blank-field
extragalactic sky at 1.1mm, to $\approx
0.4$ to 1.4~mJy beam$^{-1}$, coupled to 15 and 10m telescopes with
resolutions 18 and 30~arcsec, respectively
(Scott et al. 2008, 2010, Perera et al. 2008,
Austermann et al. 2009, 2010, Hatsukade et al. 2011, Ikarashi et
al. 2011). The COSMOS survey of 0.72~sq. deg. presented in this paper
is the largest single-field extragalactic area mapped at 1.1mm at
these depths, and
provides important constraints especially when compared to the Scuba
HAlf Degree Extragalactic Survey (SHADES) fields mapped at 1.1mm, of
comparable extension and depth (Austermann et al. 2010).

Section~2 presents the AzTEC/ASTE observations in the COSMOS field;
section~3 details the data reduction processes employed to produce our
1.1mm map; section~4 characterizes the systematic properties of the
map and derives the catalog of source candidates from our
observations; section~5 compares the AzTEC source catalog to other mm
and radio-wavelength catalogs in COSMOS; section~6 presents the number counts
over the 0.72~sq. deg. area, which shows an excess over those of the
0.5~sq deg. SHADES field; section~7 explores the origins of the excess
in terms of cross-correlations with the optical-IR galaxy population; and section~8 discusses the results and summarizes our conclusions.

\section{Observations}

We imaged a 2800~sq. arcmin field centered at right ascension
RA(J2000.0)=$10^{\rm h} 00^{\rm m} 30.00^{\rm s}$ and declination
Dec(J2000.0)= $2^{\rm o} 14' 00.0''$ with AzTEC  mounted on the
10m ASTE (Ezawa et al. 2004, 2008), located at 4800m
in the Atacama Desert of Chile. The survey was carried out from
October 20 to November 30, 2008, during excellent observing
conditions, with mean zenith opacity, as reported by the ASTE monitor,
$\tau_{220\rm GHz} =0.05$ and values of  $\tau_{220\rm GHz} < 0.06$
about 76~per cent of the time.
ASTE was operated remotely with the N-COSMOS3 network
system (Kamazaki et al. 2005) by observers deployed in San Pedro de
Atacama (Chile), Mitaka and Nobeyama (Japan), Amherst (USA) and
Tonantzintla (Mexico). A total of 112.6~hrs of on-bolometer time was
devoted to this field, excluding calibration and pointing
observations. The area was sampled in raster mode at
208~arcsec~s$^{-1}$ along 52~arcmin stripes oriented in azimuth,
spaced by 1~arcmin steps in elevation.  Since
the array orientation is fixed in azimuth and elevation, the scan
angle in RA-Dec continuously changes due to sky rotation, providing
excellent cross-linking in the final combined image. A total of 203
raster-scan maps were acquired in COSMOS during the 2008 observing
season, each taking 33 minutes of observation in a single passage
through the scanning area.  Out of the 144 nominal bolometers of AzTEC,
117 were operative during this season.

Pointing observations were acquired every half an hour to an hour,
sandwiching every one to two COSMOS raster-maps, depending on
observing conditions. The bright QSO 1055$+$018, $S_{\rm 1.1mm} \sim
2$~Jy, was used to measure shifts from the standard ASTE pointing
model. We acquired $\sim 4\times4$~arcmin$^2$ maps of our pointing
target with a continuous Lissajous pattern (Scott et al. 2010), and
fitted a bi-dimensional Gaussian to the resulting map. The calculated
offset-corrections were not implemented in real time. Instead, they
were fed into a time series, and a linearly interpolated offset was
applied to the telescope-pointing time series of each COSMOS
raster-map during the reduction process. A total of 270 pointing
observations were obtained for the COSMOS field during the campaign.

AzTEC maps are calibrated using planets as primary calibrators.  Each
night Uranus or Neptune was imaged in
order to derive the flux conversion factor for each detector.  
In a single observation of a field the typical statistical calibration
error is found to be $6-13$~per cent (Wilson et al. 2008a).  
This estimate is supported by the observations of PKS 0537$-$411, 
where we report a 6~per cent scatter in 31 single observations (Wilson et
al. 2008b). The COSMOS data presented in this paper is the weighted sum of 203 observations
taken over 27 days.  If the nightly calibration uncertainty is as high
as 13~per cent, in the absence of systematic errors, combining the 27 days of
observations leaves us with a statistical calibration uncertainty of
2.5~per cent.  Adding this in quadrature with the 5~per cent uncertainty in the
brightness temperature of Uranus at 1.1 mm (Griffin \& Orton 1993)
gives a 5.6~per cent overall calibration uncertainty.

\section{Data Reduction}

We reduced the AzTEC data in a manner similar to that described in
detail by Scott et al. (2008), but with an added set of steps to
account for nonlinearities in our atmospheric cleaning technique. For
each of our 203 observations of the COSMOS field, the raw time-stream
data from the instrument, which includes both bolometer and pointing
data, are despiked and then ``cleaned'' of atmospheric contamination
in a row-by-row manner using our standard principal component analysis
technique (see Scott et al. 2008 for a description).  An astrometric
correction is made to all pointing signals in the time stream based on
a linear interpolation of the pointing offsets measured by the
bracketing pointing observations of the QSO 1055$+$018.  With this
correction in place, the bolometer signals are flux-calibrated and
binned into $3\times 3$~arcsec$^2$ pixels.  Performing this process
for the 203 observations of the field results in 203 nearly independent
maps which are then co-added to make a preliminary image of the sky.

As in previous AzTEC analyses, we also produce 100 noise-only
realizations of the COSMOS field by jack-knifing the time-stream data
on a row-by-row basis as described in Scott et al. (2008). These noise
maps are used extensively in the characterization of the map properties.

As a deviation from previous AzTEC analyses,  we revise the technique
presented in Scott et al. (2008), used to estimate the transfer function
of the correlated noise removal algorithm on point sources in the
data. The previous technique estimated the transfer function by 
creating simulated
data that contain only a point source at the center of the field and
subtracting the eigenmodes identified for removal in the raw time-stream
data. The revised technique differs in that the transfer function is
estimated by executing the cleaning algorithm on data that combines the raw
time-stream data with several simulated point sources distributed throughout
the field. In the latter case, the eigenmodes identified for removal are
recalculated from the combination of all sources of signal, noise and
simulated point sources. In detail, several
Gaussian point sources are added to the time-stream data, all
map-making steps save optimal filtering are performed, and the
resulting maps are differenced from the original true maps of the
sky. The regions near the simulated sources are normalized, stacked
and rotationally averaged to produce the point source `kernel.' As
demonstrated in Downes et al. (2011), this technique provides
a better estimation of the effects of the non-linear correlated noise
removal on a typical point source in the map.  For the COSMOS field,
this has the effect of increasing the measured flux and noise 
values by $+15$~per
cent relative to the values obtained from the prescription in Scott et
al. (2008). We use this kernel to optimally filter the maps in an
identical manner to that in previous AzTEC analyses.

The final step in the analysis is to use the mean power spectral
density of the noise maps and the newly estimated kernel map to
construct an optimal filter for point source detection.
This final set of Wiener filtered maps is composed of the filtered
signal map, the filtered weight map, and the corresponding signal to
noise map.  The 100 noise realizations are filtered with the same
Wiener filter and will collectively be referred to as ``noise maps''.

\begin{figure}
\epsfig{file=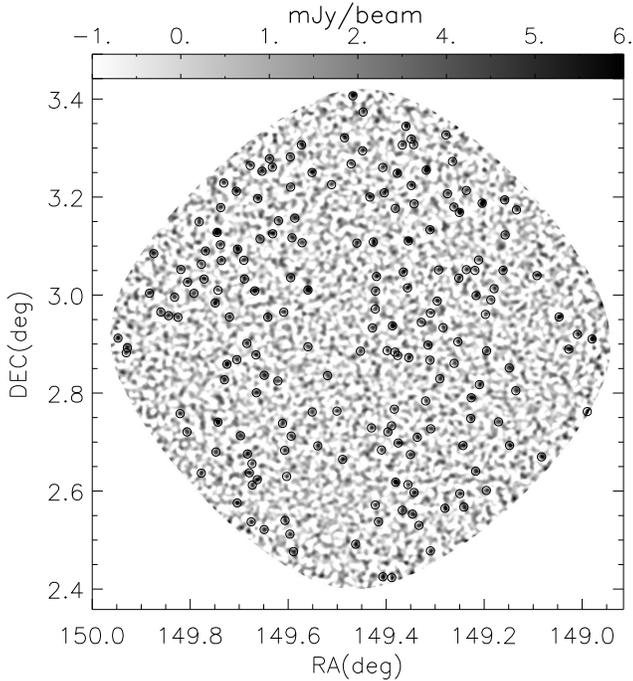,width=1.1\hsize,angle=90}
\caption{AzTEC 1.1mm map acquired at ASTE.  The circles represent
the 189 S/N$\geq 3.5$ source candidates that
have been extracted within the 50~per cent coverage
area. For reference, the circles have a radius of 30~arcsec.}
\label{fig:map08}
\end{figure}

\begin{figure}
\epsfig{file=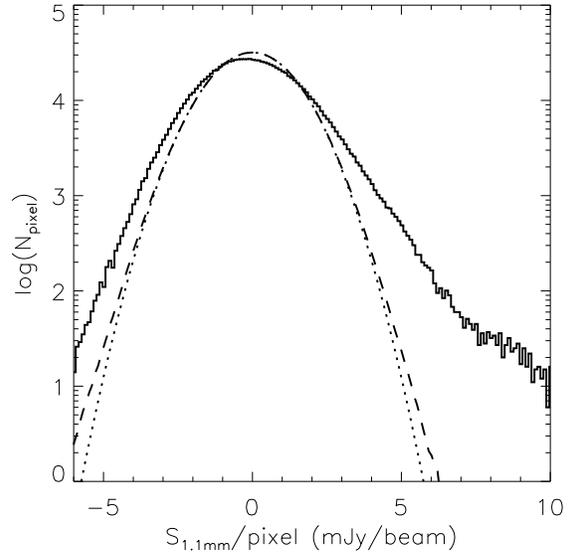,width=1.0\hsize,angle=90}
\caption{Histogram of flux density values in the AzTEC/ASTE map. The
thick solid line represents the values of the signal map over the 50~per cent
uniform
coverage area. Signal from astronomical
  sources produces both positive and negative pixels due to the fact
  that the AzTEC map has a mean of zero (Scott et al. 2008).
The  dashed line represents the average distribution of flux densities
  found in 100 jack-knifed noise realizations of the field. This
  distribution is well approximated by a Gaussian function of mean
  zero and sigma 1.26~mJy~beam$^{-1}$ (dotted line), which we adopt as
  the average noise level in the uniform coverage area of the map.}
\label{fig:FluxHisto}
\end{figure}

\section{1.1mm map and source catalogue}

\subsection{Map}

The inner 0.72~sq. deg. map of COSMOS acquired by AzTEC at ASTE is
shown in figure~\ref{fig:map08}. This area corresponds to a minimum
coverage of 50~per cent of the maximum coverage attained in the map,
and has been shown to provide excellent noise properties for source
extraction and overall population analysis (e.g. Scott et al. 2010).
The raster scan strategy for COSMOS translates into very uniform noise
properties along this section of the map, ranging from 1.23 to
1.27~mJy~beam$^{-1}$, while at the extreme edges of the map (not represented),
the noise increases quickly as one moves away from the center.  The
overall noise of the map, characterized by the combined jack-knifed
noise simulations of the individual raster-maps, is well represented
by a Gaussian distribution of rms 1.26~mJy~beam$^{-1}$ (see
Fig.~\ref{fig:FluxHisto}). This value is within the noise
range of our previous 0.15~sq. deg. COSMOS map (Scott et al. 2008).

\subsection{Astrometry}

The astrometric accuracy of the AzTEC map is verified by stacking at
the positions of 1471 radio sources in the field that are {\it not}
located within 30 arcseconds of $\pm 3.5\sigma$ or greater peaks in
the AzTEC map.  The radio source locations are taken from the Very Large Array (VLA)
1.4GHz deep mosaic of COSMOS (Schinnerer et al. 2010) which has an rms
noise level of $15 \mu$Jy~beam$^{-1}$ in the central $50'\times50'$
region, and positional accuracy better than 1~arcsec.
The stacked AzTEC map has a bright, PSF-shaped peak with
S/N$= 13$ and an offset from the center of the stacked image of
$0.6\pm1.3$~arcsec in RA and $-2.6\pm1.3$~arcsec in Dec.  Both
of these mean offset values are small compared to the 3~arcsecond
pixel size of the map and so we consider them to be too small to
warrant any correction.

\subsection{Source Catalogue}

The source extraction algorithm employed to derive the candidate
source catalog is identical to that used in Scott et al. (2008).
 We identify point sources in the 1.1mm S/N map
by searching for local maxima within 15~arcsec of pixels with S/N$\geq 3.5$
inside the 50~per cent
coverage region. The 189 source candidates are marked with circles in
figure~\ref{fig:map08} and listed in Table~\ref{tab:stacks}, together
with their measured S/N, flux densities, $1\sigma$ photometric errors,
and deboosted flux densities (see \S~\ref{sec:num_counts}). All
sources in the COSMOS catalog appear to be unresolved.

\begin{landscape}
\begin{table}
\begin{center}
\caption{AzTEC/ASTE source catalog and ancillary sub-mm to radio photometry.
The columns give (1) AzTEC identification number,
(2) source name, (3) S/N of the detection in the AzTEC map,
(4) measured 1.1mm flux density, (5) deboosted flux density and
68~per cent confidence interval, (6) flux density at $890\mu$m from SMA observations
(Younger et al. 2007, 2009),
(7) flux density at 1.2mm from MAMBO (Bertoldi et al. 2007), (8)
flux density at 1.4GHz from VLA (Schinnerer et al. 2010), (9) probability of chance association as per $P$-statistics, (10) angular distance to radio counterpart,
and (11) Notes on individual objects, including names for the source in
other (sub-)mm catalogs, where  AzTEC\_J
refers to the AzTEC/JCMT catalog (Scott et al. 2008),
AzTEC denotes interferometric SMA detections at 890$\mu$m
(Younger et al. 2007, 2009),
MMJ denotes detections at 1.2mm
by the COSBO survey performed with MAMBO
(Bertoldi et al. 2007), and B indicates
detection at 1.1mm by the Bolocam survey carried out
at the Caltech Submillimeter Observatory
(CSO; J. Aguirre, private communication; for a description of the survey see
Aguirre et al. 2006). Also included in these notes are claims
for additional radio associations (RA) and spectroscopic redshifts, where
references are (a) Younger et al. 2007; (b) Younger et al. 2009; (c) Capak et al. 2011, Riechers et al. 2010;
(d) Smol\u{c}i\'c et al. 2011; and (e) Capak et al. in preparation
}
\begin{tabular}{lcrcccccccl}
\hline
AzTEC ID &  Source name & S/N & $S_{\rm 1.1mm}$ &  $S_{\rm 1.1mm}$  & $S_{890\mu{\rm m}}$ &   $S_{\rm 1.2mm}$ & $S_{\rm 1.4GHz}$ & $P_{1.4\rm GHz}$ & $\theta$ & Notes  \\
          & (MMJ+)  &   &  (m) & (db) & & & & & & \\
  &  &   & mJy & mJy & mJy & mJy & $\mu$Jy & & $"$ & \\
\hline
AzTEC/C1 & 100141.70$+$022711.7 & 11.5 & $13.7\pm1.2$ & $13.0\pm^{1.1}_{1.0}$  & & & & & & RA$^{\rm (d)}$\\
AzTEC/C2 & 095959.20$+$023450.3 &  9.9 & $12.0\pm1.2$ & $11.2\pm^{1.1}_{1.0}$ & $19.7\pm1.8$ & & & & & AzTEC\_J095959.33$+$023445.8, AzTEC8, RA$^{\rm (b)}$\\
AzTEC/C3 & 100008.00$+$022609.7 &  9.3 & $11.3\pm1.2$ & $10.5\pm^{1.0}_{1.1}$ & $12.4\pm1.0$ & & $76.0\pm14.0$ & 0.003 &
2.5 & AzTEC\_J100008.03$+$022612.1, AzTEC2, B  \\
  &  &   &  &  & &  & & &  &      $z=1.12^{\rm (e)}$ \\
AzTEC/C4 & 095931.78$+$023047.2 &  9.2 & $11.3\pm1.2$ & $10.5\pm^{1.0}_{1.1}$ & $14.4\pm1.9$ & & & & & AzTEC\_J095931.83$+$023040.2, AzTEC4 \\
AzTEC/C5 & 095942.92$+$022939.0 &  8.9 & $10.8\pm1.2$ & $10.0\pm^{1.1}_{1.1}$ & $15.6\pm1.1$ & & & & & AzTEC\_J095942.68$+$022936.0, AzTEC1, RA$^{\rm (a)}$  \\
  &  &   &  &  & &  & & &  &   $z=4.65^{\rm (d)}$ \\
AzTEC/C6 & 100057.20$+$022008.7 &  8.7 & $10.4\pm1.2$ & $ 9.6\pm^{1.1}_{1.0}$ &
  & $7.5\pm1.1$ & $81.0\pm15.0$ & 0.007 & 3.7 & MMJ100057$+$022013, B \\
  & & & & &          & & $73.0\pm11.0$ & 0.024 & 7.0 & \\
AzTEC/C7 & 100015.77$+$021545.1 &  8.0 & $ 9.7\pm1.2$ & $ 8.9\pm^{1.1}_{1.1}$
& & $6.3\pm0.9$ &  &  &  & MMJ100016$+$021549, B\\
AzTEC/C8 & 100014.21$+$015636.1 &  7.8 & $ 9.5\pm1.2$ & $ 8.7\pm^{1.1}_{1.1}$ & & & & & &  \\
AzTEC/C9 & 100123.24$+$022002.7 &  7.5 & $ 8.9\pm1.2$ & $ 8.1\pm^{1.0}_{1.1}$ & & &  $77.0\pm16.0$ & 0.012 & 4.9 &   B\\
AzTEC/C10 & 100013.36$+$023427.2 &  7.4 & $ 9.0\pm1.2$ & $ 8.1\pm^{1.1}_{1.1}$ &  $4.4\pm1.0$ &
 & & & &  AzTEC\_J100013.21$+$023428.2, AzTEC15 \\
AzTEC/C11 & 100141.28$+$020357.1 &  7.3 & $ 8.7\pm1.2$ & $ 7.9\pm^{1.1}_{1.1}$ &  & & & & &  \\
AzTEC/C12 & 100136.87$+$021103.0 &  7.0 & $ 8.4\pm1.2$ & $ 7.5\pm^{1.0}_{1.1}$ &  & & & & &  \\
AzTEC/C13 & 095837.91$+$021408.3 &  6.8 & $ 9.9\pm1.5$ & $ 8.7\pm^{1.3}_{1.4}$ &  & & $144.0\pm13.0$ & 0.000 & 0.9 & \\
AzTEC/C14 & 095957.35$+$022732.1 &  6.3 & $ 7.7\pm1.2$ & $ 6.7\pm^{1.1}_{1.1}$ &  $9.0\pm2.2$ & & $68.0\pm13.0$ & 0.002 & 1.8 &  AzTEC\_J095957.22$+$022729.3, AzTEC9, B \\
AzTEC/C15 & 100131.67$+$022509.0 &  6.2 & $ 7.4\pm1.2$ & $ 6.5\pm^{1.1}_{1.1}$ &  & & $131.0\pm13.0$ & 0.030 & 7.9 & \\
AzTEC/C16 & 095854.11$+$021650.5 &  6.1 & $ 7.7\pm1.3$ & $ 6.7\pm^{1.1}_{1.1}$ &  & & $82.0\pm14.0$ & 0.010 & 4.5 & \\
AzTEC/C17 & 100055.19$+$023432.8 &  5.9 & $ 7.1\pm1.2$ & $ 6.2\pm^{1.1}_{1.1}$ &  & & $78.0\pm12.0$ & 0.040 & 9.2 & \\
AzTEC/C18 & 100035.19$+$024356.5 &  5.8 & $ 9.4\pm1.6$ & $ 7.9\pm^{1.4}_{1.6}$ &  $13.5\pm{1.8}$ & & $98.0\pm16.0$ & 0.006 & 3.5 &  AzTEC\_J100035.37$+$024352.3, AzTEC12 \\
AzTEC/C19 & 095950.02$+$015324.0 &  5.6 & $ 7.0\pm1.2$ & $ 5.9\pm^{1.1}_{1.1}$ &   & & & & &  \\
AzTEC/C20 & 100114.46$+$022702.5 &  5.6 & $ 6.8\pm1.2$ & $ 5.7\pm^{1.2}_{1.0}$ &   & & & & &  \\
AzTEC/C21 & 095921.55$+$022233.5 &  5.6 & $ 6.9\pm1.2$ & $ 5.9\pm^{1.0}_{1.2}$ &   & & & & &   \\
AzTEC/C22 & 100009.16$+$024012.1 &  5.6 & $ 7.5\pm1.4$ & $ 6.4\pm^{1.2}_{1.3}$ &  $4.7\pm1.3$ & & $206.0\pm15.0$ & 0.008 & 4.0 & AzTEC\_J100008.80$+$024008.0, AzTEC11 \\
  &  &   &  &  & &  & & &  &  double RA$^{\rm (b)}$    \\
AzTEC/C23 & 100142.65$+$021833.1 &  5.6 & $ 6.7\pm1.2$ & $ 5.7\pm^{1.0}_{1.1}$ &   & & $124.0\pm14.0$ & 0.013 & 5.1 & \\
     & & & & & & &  $216.0\pm15.0$ & 0.036 & 8.6 & \\
AzTEC/C24 & 100010.41$+$022220.9 &  5.6 & $ 6.8\pm1.2$ & $ 5.7\pm^{1.1}_{1.1}$ & & & $100.0\pm13.0$ & 0.008 & 4.0 & \\
AzTEC/C25 & 100122.02$+$015654.3 &  5.6 & $ 6.7\pm1.2$ & $ 5.7\pm^{1.1}_{1.1}$ &  & & $70.0\pm12.0$ & 0.050 & 10.3 & \\
AzTEC/C26 & 100132.27$+$023214.4 &  5.5 & $ 6.8\pm1.2$ & $ 5.7\pm^{1.2}_{1.1}$ &    & & & & &   \\
AzTEC/C27 & 095937.16$+$020656.5 &  5.5 & $ 6.7\pm1.2$ & $ 5.7\pm^{1.1}_{1.2}$ &     & & & & &  \\
AzTEC/C28 & 095849.35$+$021253.5 &  5.4 & $ 7.0\pm1.3$ & $ 5.9\pm^{1.2}_{1.2}$ & & & $263.0\pm18.0$ & 0.034 & 8.4 &  \\
AzTEC/C29 & 095918.38$+$020105.9 &  5.4 & $ 6.7\pm1.2$ & $ 5.6\pm^{1.1}_{1.1}$ &  & & $133.0\pm12.0$ & 0.001 & 1.2 & \\
AzTEC/C30 & 100025.15$+$022602.2 &  5.4 & $ 6.5\pm1.2$ & $ 5.5\pm^{1.1}_{1.1}$ & &  & $56.0\pm11.0$ & 0.020 & 6.5 &  AzTEC\_J100025.23$+$022608.0, B \\
AzTEC/C31 & 100147.43$+$022455.9 &  5.3 & $ 6.3\pm1.2$ & $ 5.4\pm^{1.1}_{1.1}$ &     & & & & &  \\
AzTEC/C32 & 100012.96$+$020124.1 &  5.2 & $ 6.4\pm1.2$ & $ 5.3\pm^{1.1}_{1.1}$ & & $5.2\pm2.0$  & $56.0\pm11.0$ & 0.020 & 6.5 & MMJ100012$+$020125 \\
AzTEC/C33 & 100026.82$+$023132.4 &  5.2 & $ 6.3\pm1.2$ & $ 5.3\pm^{1.1}_{1.1}$ & & & $67.0\pm12.0$ & 0.043 & 9.5 & AzTEC\_J100026.68$+$023128.1 \\\hline
\end{tabular}
\label{tab:stacks}
\end{center}
\end{table}
\end{landscape}

\setcounter{table}{0}

\begin{landscape}
\begin{table}
\begin{center}
\caption{(continuation)}
\begin{tabular}{lcrcccccccl}
\hline
AzTEC ID &  Source name & S/N & $S_{\rm 1.1mm}$ &  $S_{\rm 1.1mm}$  & $S_{890\mu{\rm m}}$ &   $S_{\rm 1.2mm}$ & $S_{\rm 1.4GHz}$ & $P_{1.4\rm GHz}$ & $\theta$ & Notes  \\
  & (MMJ+) &   &  (m) & (db) & & & & & & \\
  &  &   & mJy & mJy & mJy & mJy & $\mu$Jy & & $"$ & \\
\hline
AzTEC/C34 & 100007.77$+$021151.6 &  5.2 & $ 6.3\pm1.2$ & $ 5.3\pm^{1.1}_{1.2}$ & & $5.7\pm0.9$ & & & & MMJ100007$+$021149 \\
AzTEC/C35 & 100008.37$+$022024.3 &  5.1 & $ 6.2\pm1.2$ & $ 5.2\pm^{1.1}_{1.1}$ &   & & $62.0\pm12.0$ & 0.041 & 9.3 & \\
AzTEC/C36 & 095840.29$+$020514.7 &  5.1 & $ 8.6\pm1.7$ & $ 6.8\pm^{1.5}_{1.6}$ &  & & $168.0\pm15.0$ & 0.000 & 0.2 &  \\
AzTEC/C37 & 100121.82$+$023121.3 &  5.1 & $ 6.1\pm1.2$ & $ 5.1\pm^{1.1}_{1.1}$ &  & & $52.0\pm8.0$ & 0.031 & 8.0 & \\
AzTEC/C38 & 100023.58$+$022148.2 &  5.1 & $ 6.2\pm1.2$ & $ 5.1\pm^{1.2}_{1.1}$ &  & & $43.0\pm11.0$ & 0.025 & 7.2 & \\
AzTEC/C39 & 100126.84$+$020003.3 &  5.1 & $ 6.1\pm1.2$ & $ 5.1\pm^{1.1}_{1.1}$ &  & & & & &   \\
AzTEC/C40 & 095934.76$+$021927.6 &  5.0 & $ 6.1\pm1.2$ & $ 5.1\pm^{1.1}_{1.2}$ &  & & $64.0\pm11.0$ & 0.063 & 11.6 &  \\
AzTEC/C41 & 100148.08$+$022129.3 &  5.0 & $ 6.0\pm1.2$ & $ 4.9\pm^{1.1}_{1.1}$ &  & & & & &   \\
AzTEC/C42 & 100019.75$+$023203.4 &  4.9 & $ 5.9\pm1.2$ & $ 4.8\pm^{1.1}_{1.1}$ &  $9.3\pm{1.1}$ & & $126.0\pm15.0$ & 0.000 & 0.9 & AzTEC\_J100019.73$+$023206.0, AzTEC5, $z=3.97^{\rm (e)}$ \\
 & & & & & & & $85.0\pm15.0$ & 0.028 & 7.7 &  \\
AzTEC/C43 & 100003.58$+$020206.1 &  4.9 & $ 6.0\pm1.2$ & $ 4.8\pm^{1.2}_{1.1}$ &  & & $84.0\pm11.0$ & 0.032 & 8.2 & \\
AzTEC/C44 & 100033.80$+$014900.2 &  4.9 & $ 6.2\pm1.3$ & $ 5.0\pm^{1.2}_{1.2}$ &   & & & & &  \\
AzTEC/C45 & 100006.54$+$023257.1 &  4.9 & $ 6.0\pm1.2$ & $ 4.8\pm^{1.1}_{1.1}$ &   & & & & & \\
AzTEC/C46 & 100114.61$+$023511.9 &  4.9 & $ 6.0\pm1.2$ & $ 4.8\pm^{1.2}_{1.1}$ &  & & $122.0\pm12.0$ & 0.021 & 6.7 &  \\
AzTEC/C47 & 095941.18$+$020105.6 &  4.9 & $ 5.9\pm1.2$ & $ 4.8\pm^{1.1}_{1.1}$ & & & $222.0\pm11.0$ & 0.038 & 9.0 & \\
AzTEC/C48 & 100039.24$+$023847.9 &  4.9 & $ 6.1\pm1.2$ & $ 4.9\pm^{1.1}_{1.2}$ & & & $63.0\pm13.0$ & 0.003 & 2.5 & AzTEC\_J100038.72$+$023843.8  \\
AzTEC/C49 & 100131.83$+$015403.3 &  4.9 & $ 6.5\pm1.3$ & $ 5.3\pm^{1.2}_{1.3}$ &  & & & & &\\
AzTEC/C50 & 095933.13$+$020833.2 &  4.8 & $ 5.9\pm1.2$ & $ 4.8\pm^{1.1}_{1.2}$ &   & & $57.0\pm10.0$ & 0.001 & 1.1 &\\
AzTEC/C51 & 100040.19$+$015923.7 &  4.8 & $ 5.8\pm1.2$ & $ 4.7\pm^{1.1}_{1.1}$ & & & & & &  \\
AzTEC/C52 & 100156.23$+$022106.3 &  4.8 & $ 5.7\pm1.2$ & $ 4.7\pm^{1.1}_{1.1}$ &   & & $119.0\pm15.0$ & 0.026 & 7.3 &\\
AzTEC/C53 & 100122.65$+$021211.8 &  4.8 & $ 5.7\pm1.2$ & $ 4.6\pm^{1.1}_{1.1}$ &   & & & & &\\
AzTEC/C54 & 100125.89$+$015744.6 &  4.8 & $ 5.7\pm1.2$ & $ 4.6\pm^{1.1}_{1.1}$ &    & & & & &\\
AzTEC/C55 & 100005.19$+$015520.0 &  4.8 & $ 5.8\pm1.2$ & $ 4.7\pm^{1.1}_{1.2}$ &   & & $65.0\pm11.0$ & 0.074 & 12.5 &\\
AzTEC/C56 & 095904.93$+$022154.6 &  4.7 & $ 5.8\pm1.2$ & $ 4.7\pm^{1.1}_{1.1}$ & & & & & & \\
AzTEC/C57 & 095958.34$+$021324.4 &  4.7 & $ 5.7\pm1.2$ & $ 4.6\pm^{1.2}_{1.1}$ & & & $82.0\pm12.0$ & 0.049 & 10.2 &  \\
AzTEC/C58 & 100020.47$+$014500.6 &  4.7 & $ 7.3\pm1.6$ & $ 5.6\pm^{1.4}_{1.5}$ &  & & $59.0\pm13.0$ & 0.019 & 6.3 & \\
AzTEC/C59 & 100030.40$+$023712.2 &  4.7 & $ 5.6\pm1.2$ & $ 4.6\pm^{1.1}_{1.2}$ &  & & $161.0\pm15.0$ & 0.018 & 6.1 & \\
AzTEC/C60 & 100128.29$+$022129.7 &  4.6 & $ 5.5\pm1.2$ & $ 4.4\pm^{1.1}_{1.1}$ &  & & & & & \\
AzTEC/C61 & 100119.68$+$023442.0 &  4.6 & $ 5.7\pm1.2$ & $ 4.6\pm^{1.2}_{1.1}$ &   & & $10350\pm1000$ & 0.018 & 6.1 & extended\\
AzTEC/C62 & 100100.37$+$023756.3 &  4.6 & $ 5.9\pm1.3$ & $ 4.7\pm^{1.2}_{1.2}$ &   & & & & &\\
AzTEC/C63 & 095920.75$+$023111.3 &  4.6 & $ 6.1\pm1.3$ & $ 4.8\pm^{1.1}_{1.3}$ &   & & & & &\\
AzTEC/C64 & 100139.64$+$022345.2 &  4.6 & $ 5.5\pm1.2$ & $ 4.4\pm^{1.1}_{1.1}$ &   & & & & &\\
AzTEC/C65 & 095943.22$+$022136.1 &  4.6 & $ 5.6\pm1.2$ & $ 4.4\pm^{1.2}_{1.1}$ & & & $153.0\pm12.0$ & 0.046 & 9.8 & \\
AzTEC/C66 & 100105.00$+$022632.7 &  4.6 & $ 5.5\pm1.2$ & $ 4.3\pm^{1.1}_{1.1}$ &  & & $86.0\pm11.0$ & 0.015 & 5.5 & \\
AzTEC/C67 & 100118.61$+$020941.7 &  4.6 & $ 5.5\pm1.2$ & $ 4.3\pm^{1.1}_{1.1}$ &  & & & & & \\
AzTEC/C68 & 100122.43$+$020733.5 &  4.6 & $ 5.5\pm1.2$ & $ 4.3\pm^{1.1}_{1.1}$ &   & & & & &\\
AzTEC/C69 & 100138.07$+$020908.9 &  4.6 & $ 5.5\pm1.2$ & $ 4.3\pm^{1.1}_{1.1}$ &   & & & & &\\
AzTEC/C70 & 100026.04$+$020314.9 &  4.5 & $ 5.5\pm1.2$ & $ 4.3\pm^{1.1}_{1.1}$ &  & & $547.0\pm13.0$ & 0.022 & 6.8 & \\
         & & & & & & & $49.0\pm13.0$ & 0.037 & 8.8 & \\
AzTEC/C71 & 095953.83$+$021847.9 &  4.5 & $ 5.5\pm1.2$ & $ 4.3\pm^{1.1}_{1.1}$ & &  $5.7\pm$  1.3& $79.0\pm11.0$ & 0.017 & 6.0 & MMJ095953$+$021851, B \\
AzTEC/C72 & 100159.82$+$020459.8 &  4.5 & $ 5.6\pm1.2$ & $ 4.4\pm^{1.2}_{1.2}$ & & & & & &  \\
AzTEC/C73 & 100103.62$+$022856.8 &  4.5 & $ 5.4\pm1.2$ & $ 4.2\pm^{1.1}_{1.1}$ & & & & & &  \\
AzTEC/C74 & 100105.64$+$022139.6 &  4.5 & $ 5.4\pm1.2$ & $ 4.2\pm^{1.1}_{1.1}$ &  & & $91.0\pm12.0$ & 0.025 & 7.3 &  \\
\hline
\end{tabular}
\label{tab:stacks}
\end{center}
\end{table}
\end{landscape}

\setcounter{table}{0}

\begin{landscape}
\begin{table}
\begin{center}
\caption{(continuation)}
\begin{tabular}{lcrcccccccl}
\hline
AzTEC ID &  Source name & S/N & $S_{\rm 1.1mm}$ &  $S_{\rm 1.1mm}$  & $S_{890\mu{\rm m}}$ &   $S_{\rm 1.2mm}$ & $S_{\rm 1.4GHz}$ & $P_{1.4\rm GHz}$ & $\theta$ & Notes  \\
  & (MMJ+)  &   &  (m) & (db) & & & & & & \\
  &  &   & mJy & mJy & mJy & mJy & $\mu$Jy & & $"$ & \\
\hline
AzTEC/C75 & 100052.35$+$020103.1 &  4.4 & $ 5.3\pm1.2$ & $ 4.2\pm^{1.1}_{1.2}$ & & & & & &  \\
AzTEC/C76 & 100013.21$+$021207.0 &  4.4 & $ 5.4\pm1.2$ & $ 4.2\pm^{1.1}_{1.1}$ & & & & & &  \\
AzTEC/C77 & 095935.00$+$015756.7 &  4.4 & $ 5.4\pm1.2$ & $ 4.2\pm^{1.2}_{1.1}$ &  & & $69.0\pm11.0$ & 0.014 & 5.4 &  \\
AzTEC/C78 & 095902.73$+$015942.0 &  4.4 & $ 6.5\pm1.5$ & $ 4.8\pm^{1.4}_{1.3}$ & & & & & &   \\
AzTEC/C79 & 095943.70$+$021348.3 &  4.4 & $ 5.4\pm1.2$ & $ 4.2\pm^{1.1}_{1.2}$ & & & & & &   \\
AzTEC/C80 & 100033.24$+$022553.6 &  4.4 & $ 5.3\pm1.2$ & $ 4.1\pm^{1.1}_{1.1}$ & & & & & &   \\
AzTEC/C81 & 100005.97$+$015241.4 &  4.4 & $ 5.4\pm1.2$ & $ 4.1\pm^{1.2}_{1.1}$ & & & & & &   \\
AzTEC/C82 & 100116.85$+$021648.4 &  4.4 & $ 5.2\pm1.2$ & $ 4.1\pm^{1.0}_{1.2}$ & & & & & &   \\
AzTEC/C83 & 100230.25$+$021414.6 &  4.3 & $ 6.8\pm1.6$ & $ 4.9\pm^{1.5}_{1.5}$ & & & & & &   \\
AzTEC/C84 & 095942.76$+$015511.7 &  4.3 & $ 5.3\pm1.2$ & $ 4.1\pm^{1.1}_{1.2}$ & & & & & &   \\
AzTEC/C85 & 100140.09$+$022541.3 &  4.3 & $ 5.2\pm1.2$ & $ 4.0\pm^{1.1}_{1.1}$ &  & & $549.0\pm12.0$ & 0.039 & 9.0 & \\
AzTEC/C86 & 100109.03$+$021726.0 &  4.3 & $ 5.2\pm1.2$ & $ 4.0\pm^{1.1}_{1.1}$ &  & & $4210\pm400$ & 0.016 & 5.7 & extended \\
AzTEC/C87 & 100205.50$+$021700.1 &  4.3 & $ 5.2\pm1.2$ & $ 4.0\pm^{1.1}_{1.2}$ &  & & & & & \\
AzTEC/C88 & 095937.37$+$020423.9 &  4.3 & $ 5.3\pm1.2$ & $ 4.0\pm^{1.2}_{1.1}$ &  & & & & & \\
AzTEC/C89 & 100127.08$+$021336.1 &  4.3 & $ 5.1\pm1.2$ & $ 4.0\pm^{1.1}_{1.1}$ &  & & & & & \\
AzTEC/C90 & 100135.64$+$021650.2 &  4.3 & $ 5.1\pm1.2$ & $ 4.0\pm^{1.1}_{1.2}$ &  & & & & & \\
AzTEC/C91 & 100128.61$+$022347.4 &  4.2 & $ 5.1\pm1.2$ & $ 3.8\pm^{1.2}_{1.1}$ &  & & $81.0\pm12.0$ & 0.038 & 8.9 & \\
AzTEC/C92 & 100139.94$+$023015.0 &  4.2 & $ 5.2\pm1.2$ & $ 4.0\pm^{1.1}_{1.2}$ &   & & & & & \\
AzTEC/C93 & 100132.02$+$021137.0 &  4.2 & $ 5.1\pm1.2$ & $ 3.8\pm^{1.1}_{1.1}$ &  & & $60.0\pm11.0$ & 0.004 & 2.8 &\\
AzTEC/C94 & 095957.13$+$021719.0 &  4.2 & $ 5.1\pm1.2$ & $ 3.9\pm^{1.1}_{1.2}$ &   & & & & & \\
AzTEC/C95 & 100018.36$+$021242.9 &  4.2 & $ 5.1\pm1.2$ & $ 3.8\pm^{1.2}_{1.1}$ &   & & & & & \\
AzTEC/C96 & 100108.38$+$015154.6 &  4.2 & $ 5.2\pm1.2$ & $ 3.9\pm^{1.2}_{1.2}$ &   & & & & & \\
AzTEC/C97 & 100214.84$+$021944.9 &  4.2 & $ 5.5\pm1.3$ & $ 4.1\pm^{1.2}_{1.3}$ &  & & $55.0\pm16.0$ & 0.013 & 5.1 &\\
AzTEC/C98 & 100042.99$+$020518.4 &  4.2 & $ 5.0\pm1.2$ & $ 3.8\pm^{1.1}_{1.2}$ &  & & $78.0\pm14.0$ & 0.005 & 3.1 & \\
AzTEC/C99 & 100006.98$+$015958.9 &  4.2 & $ 5.1\pm1.2$ & $ 3.8\pm^{1.1}_{1.2}$ &   & & & & & \\
AzTEC/C100 & 095918.52$+$021035.8 &  4.2 & $ 5.1\pm1.2$ & $ 3.8\pm^{1.2}_{1.1}$ &   & & & & & \\
AzTEC/C101 & 095945.58$+$023018.5 &  4.2 & $ 5.1\pm1.2$ & $ 3.8\pm^{1.1}_{1.2}$ &   & & & & & \\
AzTEC/C102 & 095948.77$+$023156.5 &  4.1 & $ 5.1\pm1.2$ & $ 3.8\pm^{1.1}_{1.2}$ &   & & & & & \\
AzTEC/C103 & 100124.43$+$015615.1 &  4.1 & $ 5.0\pm1.2$ & $ 3.8\pm^{1.1}_{1.2}$ &  & & $51.0\pm15.0$ & 0.036 & 8.7 & \\
AzTEC/C104 & 095940.74$+$015333.1 &  4.1 & $ 5.3\pm1.3$ & $ 4.0\pm^{1.2}_{1.2}$ &  & & & & &  \\
AzTEC/C105 & 095845.11$+$021442.1 &  4.1 & $ 5.5\pm1.3$ & $ 4.1\pm^{1.2}_{1.3}$ &   & & & & & \\
AzTEC/C106 & 100006.54$+$023838.6 &  4.1 & $ 5.3\pm1.3$ & $ 4.0\pm^{1.2}_{1.3}$ &  $8.6\pm1.3$ & & & & & AzTEC\_J100006.40$+$023839.8, AzTEC6 \\
AzTEC/C107 & 095939.56$+$022238.3 &  4.1 & $ 5.0\pm1.2$ & $ 3.8\pm^{1.1}_{1.2}$ &  & & & & & \\
AzTEC/C108 & 100116.05$+$023614.3 &  4.1 & $ 5.3\pm1.3$ & $ 4.0\pm^{1.2}_{1.3}$ &   & & & & &\\
AzTEC/C109 & 100111.63$+$022838.3 &  4.1 & $ 5.0\pm1.2$ & $ 3.7\pm^{1.1}_{1.1}$ &   & & $59.0\pm11.0$ & 0.004 & 2.7 &\\
AzTEC/C110 & 100108.56$+$020029.7 &  4.1 & $ 5.0\pm1.2$ & $ 3.7\pm^{1.1}_{1.1}$ &  & & & & &\\
AzTEC/C111 & 095929.62$+$021241.6 &  4.1 & $ 5.0\pm1.2$ & $ 3.7\pm^{1.2}_{1.2}$ &  & & $67.0\pm12.0$ & 0.019 & 6.3 & \\
AzTEC/C112 & 100010.94$+$015309.3 &  4.1 & $ 5.0\pm1.2$ & $ 3.7\pm^{1.1}_{1.2}$ & & & $122.0\pm12.0$ & 0.014 & 5.3 &  \\
AzTEC/C113 & 095914.96$+$022957.7 &  4.1 & $ 5.5\pm1.3$ & $ 4.0\pm^{1.3}_{1.3}$ &   & & $173.0\pm16.0$ & 0.036 & 8.7 &\\
AzTEC/C114 & 100024.06$+$022000.6 &  4.1 & $ 4.9\pm1.2$ & $ 3.7\pm^{1.1}_{1.2}$ & & & & & & \\
AzTEC/C115 & 100014.84$+$020532.5 &  4.1 & $ 4.9\pm1.2$ & $ 3.7\pm^{1.1}_{1.2}$ & & & & & & \\
\hline
\end{tabular}
\end{center}
\end{table}
\end{landscape}

\setcounter{table}{0}
\begin{landscape}
\begin{table}
\begin{center}
\caption{(continuation)}
\begin{tabular}{lcrcccccccl}
\hline
AzTEC ID &  Source name & S/N & $S_{\rm 1.1mm}$ &  $S_{\rm 1.1mm}$  & $S_{890\mu{\rm m}}$ &   $S_{\rm 1.2mm}$ & $S_{\rm 1.4GHz}$ & $P_{1.4\rm GHz}$ & $\theta$ & Notes  \\
 & (MMJ+) &   &  (m) & (db) & & & & & & \\
  &  &   & mJy & mJy & mJy & mJy & $\mu$Jy & & $"$ & \\
\hline
AzTEC/C116 & 100109.63$+$020348.3 &  4.0 & $ 4.9\pm1.2$ & $ 3.7\pm^{1.1}_{1.2}$ & & $4.2\pm1.4$ & $59.0\pm11.0$ & 0.008 & 4.0 & MMJ100109$+$020346  \\
AzTEC/C117 & 095925.93$+$022018.3 &  4.0 & $ 5.0\pm1.2$ & $ 3.7\pm^{1.1}_{1.2}$ & & & & & &  \\
AzTEC/C118 & 095959.49$+$020633.1 &  4.0 & $ 4.9\pm1.2$ & $ 3.7\pm^{1.1}_{1.2}$ & & $5.6\pm1.1$ & $104.0\pm13.0$ & 0.023 & 6.9 & MMJ100000$+$020634, B \\
AzTEC/C119 & 095915.34$+$020748.3 &  4.0 & $ 5.0\pm1.2$ & $ 3.7\pm^{1.1}_{1.2}$ & & & & & & \\
AzTEC/C120 & 100105.38$+$020214.6 &  4.0 & $ 4.9\pm1.2$ & $ 3.6\pm^{1.2}_{1.1}$ & & & & & & \\
AzTEC/C121 & 095952.53$+$020915.9 &  4.0 & $ 4.9\pm1.2$ & $ 3.7\pm^{1.1}_{1.2}$ & & & & & &\\
AzTEC/C122 & 100200.88$+$021648.0 &  4.0 & $ 4.8\pm1.2$ & $ 3.6\pm^{1.1}_{1.2}$ & & & & & & \\
AzTEC/C123 & 100022.62$+$015145.4 &  4.0 & $ 4.9\pm1.2$ & $ 3.7\pm^{1.1}_{1.2}$ & & & & & & \\
AzTEC/C124 & 095946.32$+$023554.0 &  4.0 & $ 5.1\pm1.3$ & $ 3.7\pm^{1.2}_{1.2}$ &  & & $27.0\pm13.0$ & 0.038 & 8.9 &  AzTEC\_J095946.66$+$023541.9 \\
 & & & & & & & $18059\pm1800$ & 0.033 & 8.3 & extended  \\
AzTEC/C125 & 095920.54$+$022653.8 &  4.0 & $ 4.9\pm1.2$ & $ 3.6\pm^{1.2}_{1.2}$ & & & & & & \\
AzTEC/C126 & 100159.48$+$022239.4 &  4.0 & $ 4.8\pm1.2$ & $ 3.5\pm^{1.2}_{1.2}$ &  & & $134.0\pm13.0$ & 0.048 & 10.1 & \\
AzTEC/C127 & 100125.46$+$023524.3 &  4.0 & $ 5.4\pm1.3$ & $ 3.8\pm^{1.3}_{1.3}$ &  & & $131.0\pm12.0$ & 0.007 & 3.9 &  \\
AzTEC/C128 & 100002.80$+$015118.3 &  4.0 & $ 4.9\pm1.2$ & $ 3.6\pm^{1.2}_{1.2}$ &  & & & & & \\
AzTEC/C129 & 100130.23$+$020217.6 &  4.0 & $ 4.7\pm1.2$ & $ 3.5\pm^{1.1}_{1.1}$ & & & & & & B \\
AzTEC/C130 & 095956.99$+$020308.4 &  3.9 & $ 4.8\pm1.2$ & $ 3.5\pm^{1.1}_{1.2}$ &  & & $124.0\pm10.0$ & 0.018 & 6.1 & \\
AzTEC/C131 & 100010.93$+$023754.3 &  3.9 & $ 4.9\pm1.2$ & $ 3.5\pm^{1.2}_{1.2}$ & & & & & &  \\
AzTEC/C132 & 100005.35$+$023757.4 &  3.9 & $ 4.9\pm1.3$ & $ 3.5\pm^{1.2}_{1.2}$ & & & & & &  \\
AzTEC/C133 & 100120.64$+$022624.4 &  3.9 & $ 4.7\pm1.2$ & $ 3.4\pm^{1.1}_{1.2}$ &  & & $79.0\pm12.0$ & 0.006 & 3.4 & \\
AzTEC/C134 & 100106.01$+$015014.2 &  3.9 & $ 5.0\pm1.3$ & $ 3.6\pm^{1.2}_{1.3}$ & & & & & &   \\
AzTEC/C135 & 100024.18$+$015348.4 &  3.9 & $ 4.7\pm1.2$ & $ 3.4\pm^{1.2}_{1.2}$ & & & & & &   \\
AzTEC/C136 & 095933.52$+$022348.9 &  3.9 & $ 4.7\pm1.2$ & $ 3.4\pm^{1.2}_{1.2}$ & & & $76.0\pm14.0$ & 0.001 & 1.6 & \\
AzTEC/C137 & 095953.11$+$022236.1 &  3.9 & $ 4.7\pm1.2$ & $ 3.4\pm^{1.2}_{1.2}$ & & & & & &  \\
AzTEC/C138 & 100020.54$+$023509.3 &  3.9 & $ 4.7\pm1.2$ & $ 3.4\pm^{1.1}_{1.2}$ & $8.7\pm1.5$ &  &  & & &  AzTEC\_J100020.71$+$023518.2, AzTEC3, $z=5.3^{\rm (c)}$  \\
AzTEC/C139 & 100202.55$+$021915.3 &  3.9 & $ 4.7\pm1.2$ & $ 3.4\pm^{1.1}_{1.2}$ &  & & & & &  \\
AzTEC/C140 & 100124.98$+$015144.4 &  3.9 & $ 5.4\pm1.4$ & $ 3.7\pm^{1.3}_{1.4}$ &  & & $49.0\pm11.0$ & 0.004 & 2.9 &\\
AzTEC/C141 & 100209.29$+$021727.3 &  3.9 & $ 4.7\pm1.2$ & $ 3.4\pm^{1.1}_{1.2}$ &  & & & & & \\
AzTEC/C142 & 100018.01$+$020245.5 &  3.8 & $ 4.7\pm1.2$ & $ 3.3\pm^{1.2}_{1.2}$ &  & & & & & \\
AzTEC/C143 & 100149.44$+$015742.3 &  3.8 & $ 5.4\pm1.4$ & $ 3.7\pm^{1.3}_{1.4}$ &  & & & & & \\
AzTEC/C144 & 100142.27$+$020017.7 &  3.8 & $ 4.6\pm1.2$ & $ 3.3\pm^{1.1}_{1.2}$ &  & & $247.0\pm12.0$ & 0.012 & 5.1 & \\
AzTEC/C145 & 100031.47$+$021239.0 &  3.8 & $ 4.6\pm1.2$ & $ 3.3\pm^{1.1}_{1.2}$ &  & $5.3\pm0.9$ & $189.0\pm10.0$ & 0.023 & 6.9 &MMJ100031$+$021241 \\
AzTEC/C146 & 095957.15$+$014811.0 &  3.8 & $ 5.6\pm1.5$ & $ 3.7\pm^{1.5}_{1.4}$ &  & & $77.0\pm13.0$ & 0.025 & 7.2 &\\
AzTEC/C147 & 100107.60$+$015718.3 &  3.8 & $ 4.6\pm1.2$ & $ 3.2\pm^{1.2}_{1.2}$ &  & & & & & \\
AzTEC/C148 & 100025.56$+$021530.1 &  3.8 & $ 4.6\pm1.2$ & $ 3.3\pm^{1.2}_{1.2}$ &  & $4.9\pm0.9$ & & & &MMJ100026$+$021529 \\
AzTEC/C149 & 100150.50$+$022829.7 &  3.8 & $ 4.9\pm1.3$ & $ 3.4\pm^{1.2}_{1.3}$ &  & & $95.0\pm11.0$ & 0.026 & 7.3 & \\
AzTEC/C150 & 100005.15$+$023042.3 &  3.8 & $ 4.6\pm1.2$ & $ 3.3\pm^{1.2}_{1.2}$ & & & $96.0\pm15.0$ & 0.016 & 5.8 & AzTEC\_J100004.54$+$023040.1 \\
AzTEC/C151 & 095950.96$+$021532.3 &  3.8 & $ 4.6\pm1.2$ & $ 3.2\pm^{1.2}_{1.2}$ &  & & & & & \\
AzTEC/C152 & 100030.16$+$024157.3 &  3.8 & $ 5.3\pm1.4$ & $ 3.6\pm^{1.4}_{1.4}$ &  & & & & & \\
AzTEC/C153 & 100212.74$+$022436.2 &  3.8 & $ 5.7\pm1.5$ & $ 3.7\pm^{1.5}_{1.5}$ &   & & & & &\\
AzTEC/C154 & 100036.01$+$023536.0 &  3.8 & $ 4.6\pm1.2$ & $ 3.2\pm^{1.2}_{1.2}$ &  & & $193.0\pm15.0$ & 0.032 & 8.2 &  \\
 & & & & & & &  $72.0\pm14.0$ & 0.046 & 9.8 & \\
AzTEC/C155 & 100055.08$+$020511.8 &  3.8 & $ 4.5\pm1.2$ & $ 3.2\pm^{1.1}_{1.2}$ &  & & & & & \\
AzTEC/C156 & 100225.69$+$021303.8 &  3.8 & $ 5.5\pm1.5$ & $ 3.7\pm^{1.4}_{1.5}$ &  & & & & & \\
\hline
\end{tabular}
\end{center}
\end{table}
\end{landscape}

\setcounter{table}{0}
\begin{landscape}
\begin{table}
\begin{center}
\caption{(continuation)}
\begin{tabular}{lcrcccccccl}
\hline
AzTEC ID &  Source name & S/N & $S_{\rm 1.1mm}$ &  $S_{\rm 1.1mm}$  & $S_{890\mu{\rm m}}$ &   $S_{\rm 1.2mm}$ & $S_{\rm 1.4GHz}$ & $P_{1.4\rm GHz}$ & $\theta$ & Notes  \\
  & (MMJ+) &   &  (m) & (db) & & & & & & \\
  &  &   & mJy & mJy & mJy & mJy & $\mu$Jy & & $"$ & \\
\hline
AzTEC/C157 & 100104.22$+$014805.5 &  3.8 & $ 5.3\pm1.4$ & $ 3.6\pm^{1.4}_{1.4}$ &   & & $114.0\pm12.0$ & 0.059 & 11.1 &\\
AzTEC/C158 & 100014.58$+$021232.3 &  3.7 & $ 4.5\pm1.2$ & $ 3.2\pm^{1.1}_{1.2}$ & & & & & &  \\
AzTEC/C159 & 095929.75$+$015535.4 &  3.7 & $ 4.8\pm1.3$ & $ 3.3\pm^{1.3}_{1.3}$ &  & & & & & \\
AzTEC/C160 & 100024.20$+$021748.7 &  3.7 & $ 4.5\pm1.2$ & $ 3.1\pm^{1.2}_{1.2}$ &  & $5.2\pm0.9$ & $83.0\pm12.0$ & 0.005 & 3.3 & MMJ100024$+$021748 \\
AzTEC/C161 & 095939.54$+$023220.5 &  3.7 & $ 4.5\pm1.2$ & $ 3.1\pm^{1.2}_{1.2}$ &  & $5.2\pm0.9$ & $83.0\pm12.0$ & 0.005 & 3.3 & \\
AzTEC/C162 & 100001.59$+$021611.2 &  3.7 & $ 4.5\pm1.2$ & $ 3.1\pm^{1.2}_{1.2}$ &  & & & & &  \\
AzTEC/C163 & 100124.60$+$015851.8 &  3.7 & $ 4.4\pm1.2$ & $ 3.1\pm^{1.1}_{1.2}$ &   & & & & & \\
AzTEC/C164 & 095935.35$+$022232.8 &  3.7 & $ 4.5\pm1.2$ & $ 3.1\pm^{1.2}_{1.2}$ &   & & & & & \\
AzTEC/C165 & 095949.52$+$023908.9 &  3.7 & $ 5.4\pm1.5$ & $ 3.5\pm^{1.4}_{1.5}$ &   & & & & & \\
AzTEC/C166 & 100105.81$+$023627.3 &  3.7 & $ 4.5\pm1.2$ & $ 3.1\pm^{1.2}_{1.2}$ &   & & & & & \\
AzTEC/C167 & 095923.81$+$020359.5 &  3.7 & $ 4.5\pm1.2$ & $ 3.1\pm^{1.2}_{1.3}$ &   & & & & & \\
AzTEC/C168 & 100057.18$+$021311.7 &  3.7 & $ 4.4\pm1.2$ & $ 3.0\pm^{1.2}_{1.2}$ &  & & & & & MMJ100057$+$021305 \\
AzTEC/C169 & 100014.35$+$023009.0 &  3.7 & $ 4.4\pm1.2$ & $ 3.1\pm^{1.1}_{1.2}$ &  & & & & &  \\
AzTEC/C170 & 100045.62$+$023303.6 &  3.7 & $ 4.4\pm1.2$ & $ 3.0\pm^{1.2}_{1.2}$ &   & & & & & \\
AzTEC/C171 & 100047.57$+$020939.4 &  3.6 & $ 4.4\pm1.2$ & $ 3.0\pm^{1.2}_{1.2}$ &  & & $362.0\pm13.0$ & 0.002 & 1.9 &  \\
AzTEC/C172 & 100021.15$+$020032.3 &  3.6 & $ 4.4\pm1.2$ & $ 3.0\pm^{1.2}_{1.2}$ &  & & & & & \\
AzTEC/C173 & 100118.61$+$015047.9 &  3.6 & $ 5.0\pm1.4$ & $ 3.3\pm^{1.4}_{1.4}$ &  & & & & & \\
AzTEC/C174 & 100008.19$+$015618.1 &  3.6 & $ 4.4\pm1.2$ & $ 3.0\pm^{1.2}_{1.2}$ &   & & $50.0\pm9.0$ & 0.017 & 5.9 &\\
AzTEC/C175 & 100156.44$+$020245.4 &  3.6 & $ 4.6\pm1.3$ & $ 3.1\pm^{1.2}_{1.3}$ &  & & & & &\\
AzTEC/C176 & 095957.35$+$021133.2 &  3.6 & $ 4.4\pm1.2$ & $ 3.0\pm^{1.2}_{1.2}$ & & $6.0\pm1.1$ & $60.0\pm11.0$ & 0.013 & 5.2 &MMJ095956$+$021139 \\
AzTEC/C177 & 100138.47$+$023315.4 &  3.6 & $ 5.1\pm1.4$ & $ 3.2\pm^{1.4}_{1.4}$ &  & & & & &\\
AzTEC/C178 & 100141.28$+$022006.6 &  3.6 & $ 4.3\pm1.2$ & $ 2.9\pm^{1.2}_{1.2}$ &  & & $83.0\pm13.0$ & 0.010 & 4.5 &\\
AzTEC/C179 & 100149.44$+$022318.4 &  3.6 & $ 4.3\pm1.2$ & $ 2.9\pm^{1.2}_{1.2}$ &  & & & & &\\
AzTEC/C180 & 095927.55$+$021856.5 &  3.6 & $ 4.4\pm1.2$ & $ 3.0\pm^{1.1}_{1.3}$ &  & & & & &\\
AzTEC/C181 & 095930.18$+$021709.7 &  3.6 & $ 4.4\pm1.2$ & $ 2.9\pm^{1.2}_{1.2}$ &  & & $54.0\pm10.0$ & 0.047 & 10.0 & \\
AzTEC/C182 & 100016.21$+$020329.7 &  3.6 & $ 4.3\pm1.2$ & $ 2.9\pm^{1.2}_{1.2}$ &   & & & & &\\
AzTEC/C183 & 100226.23$+$021227.1 &  3.6 & $ 5.3\pm1.5$ & $ 3.3\pm^{1.4}_{1.6}$ &  & & $71.0\pm15.0$ & 0.000 & 0.8 &\\
 & & & & & & & $83.0\pm16.0$ & 0.003 & 2.6 & \\
AzTEC/C184 & 100016.14$+$014454.3 &  3.6 & $ 5.8\pm1.6$ & $ 3.5\pm^{1.6}_{1.8}$ &  & & $93.0\pm14.0$ & 0.054 & 10.7 & \\
AzTEC/C185 & 100111.89$+$020859.8 &  3.6 & $ 4.3\pm1.2$ & $ 2.9\pm^{1.1}_{1.3}$ & & & & & &B \\
AzTEC/C186 & 100105.68$+$023245.3 &  3.5 & $ 4.3\pm1.2$ & $ 2.9\pm^{1.1}_{1.2}$ &  & & $82.0\pm15.0$ & 0.025 & 7.2 &  \\
AzTEC/C187 & 100153.08$+$021944.0 &  3.5 & $ 4.2\pm1.2$ & $ 2.9\pm^{1.1}_{1.3}$ &  & & & & & \\
AzTEC/C188 & 095945.52$+$021109.7 &  3.5 & $ 4.3\pm1.2$ & $ 2.9\pm^{1.2}_{1.3}$ &   & & & & &\\
AzTEC/C189 & 100100.06$+$022556.3 &  3.5 & $ 4.2\pm1.2$ & $ 2.8\pm^{1.2}_{1.2}$ &  & & & & & \\
\end{tabular}
\end{center}
\end{table}
\end{landscape}

\normalsize

\subsection{False detection rate}

\begin{figure}
\epsfig{file=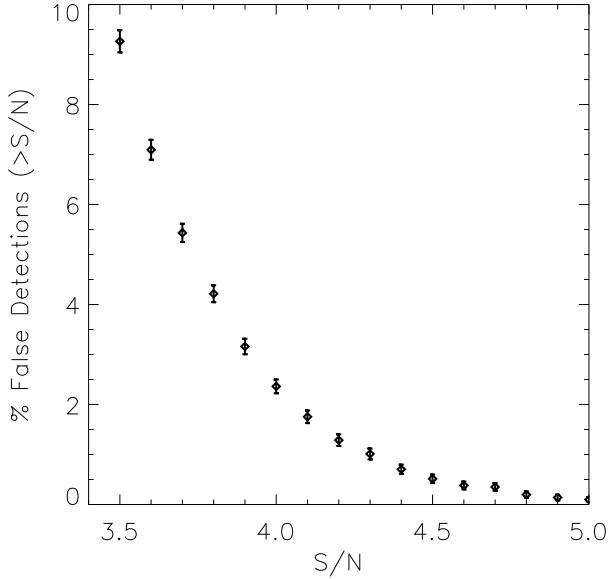,width=1.05\hsize,angle=90}
\caption{Expected fraction of false detections in the AzTEC catalog
as a function of limiting S/N estimated through counting peaks in
100 noise-only realizations of the field.
}
\label{fig:FDR}
\end{figure}

The nature of false-positives in source identification in AzTEC maps
is richly discussed in Perera et al. (2008) and Scott et al. (2010).
We follow the same procedure as Scott et al. (2010) and search our 100
noise realizations of the COSMOS field in order to derive a
conservative upper limit to the fraction of sources in
Table~\ref{tab:stacks} that are actually noise peaks rather than real
sources.  Figure~\ref{fig:FDR} shows this upper limit of the fraction of
false detections as a function of S/N.  The catalog is robust.  At
S/N$\geq 3.5$ we derive a mean of 17.4 false detections (9~per cent of
the source candidates), and at S/N$\geq 4.0$ we expect only 3 (2~per
cent) to be false.

\subsection{Completeness}

\begin{figure}
\epsfig{file=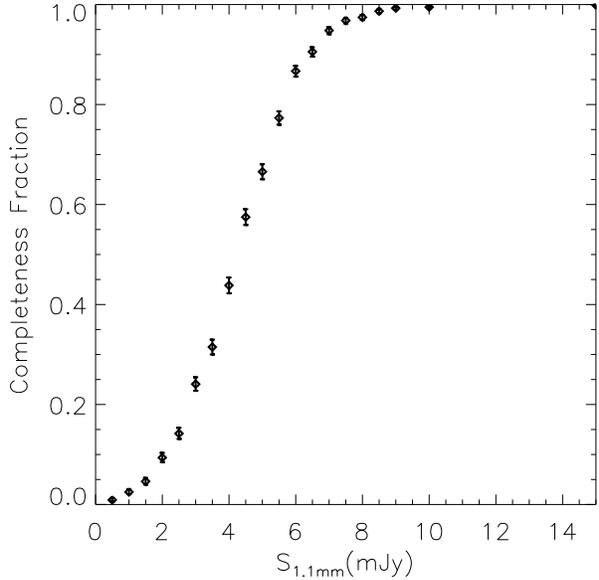,width=1.05\hsize,angle=90}
\caption{Survey completeness estimates for the COSMOS catalog
as a function of flux density. The
points and 68~per cent binomial error bars show the completeness
estimated by inserting sources of known flux density one at a time
into the observed signal map and then finding them with the same
source extraction algorithm as that used to create the
candidate source catalog. In order for a source to be considered
recovered, it must be detected with S/N$\geq 3.5$ at a distance
$r\leq 17"$ from the input location.  }
\label{fig:comp}
\end{figure}

We estimate the completeness of source detections through a set of
simulations in which artificial sources of different flux
densities are inserted
one at a time within the 50~per cent coverage of the observed COSMOS
map, and then retrieved with the same source extraction algorithm as
that used for building the catalog. As described in detail by Scott et
al. (2008, 2010) this method has the advantage of taking into account
the effects of random and confusion noise in the signal map while it does
not alter the properties of the real map significantly. The
simulations are based on 1000 test sources per represented flux value.
A test source is considered recovered if it
is extracted with S/N$\geq 3.5$ at a radius $\leq
17$~arcsec of its input position. This radius is adopted to ensure a $\sim 100$~per
cent recovery of
S/N$=3.5$ sources (see \S4.6).
Figure~\ref{fig:comp} summarizes our findings. Sources with flux
densities $S_{1.1\rm mm}\geq 6$~mJy are identified in this survey
with  $\geq 90$~per cent completeness.

\begin{figure}
\epsfig{file=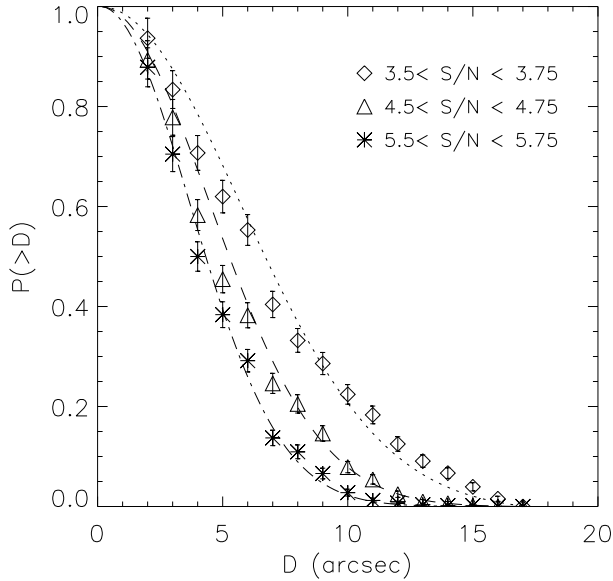,width=1.05\hsize,angle=90}
\caption{Positional uncertainty distributions for AzTEC/ASTE
source candidates. The data-points and 68~per cent confidence intervals
show the probability $P(>D)$
that AzTEC sources of different S/N will be found outside
a radial distance $D$, as determined from simulations. The
curves show the expected probabilities from the simple approximation that
takes into a account the beam size and the S/N of the detection
(Ivison et al. 2007).}
\label{fig:astrom}
\end{figure}

\subsection{Positional uncertainty}
\label{sec:posuncertainty}

The large beam of ASTE combined with low S/N detections and confusion
noise contribute to introducing random offsets in the position in
which a source is found in a map. We characterize the positional
uncertainty as a function of S/N from the same set of simulations as
those used for the completeness calculation, where now we focus on the
distance at which the sources get recovered. Figure~\ref{fig:astrom}
shows the uncertainty distribution for sources found in three S/N
regimes, and the comparison with the analytical approximation of
uncertainties (Ivison et al. 2007),
assuming a FWHM of 34~arcsec for the effective beam of
the map:
the probability
of finding a source at distance $>D$ from its true position is given by
$P(>D)=1-\exp(-D^2/2\sigma^2) $ , where $\sigma=\Delta \alpha = \Delta
\delta \approx 0.6 \, {\rm FWHM} \, ({\rm S/N})^{-1}$.
 The analytical form shows the same trend
as the empirical distribution, and can be used to estimate search
radii of possible counterparts of the SMGs presented in this paper.

\section{Comparison with other maps}

\subsection{Overlap with other mm surveys}

\begin{figure*}
\epsfig{file=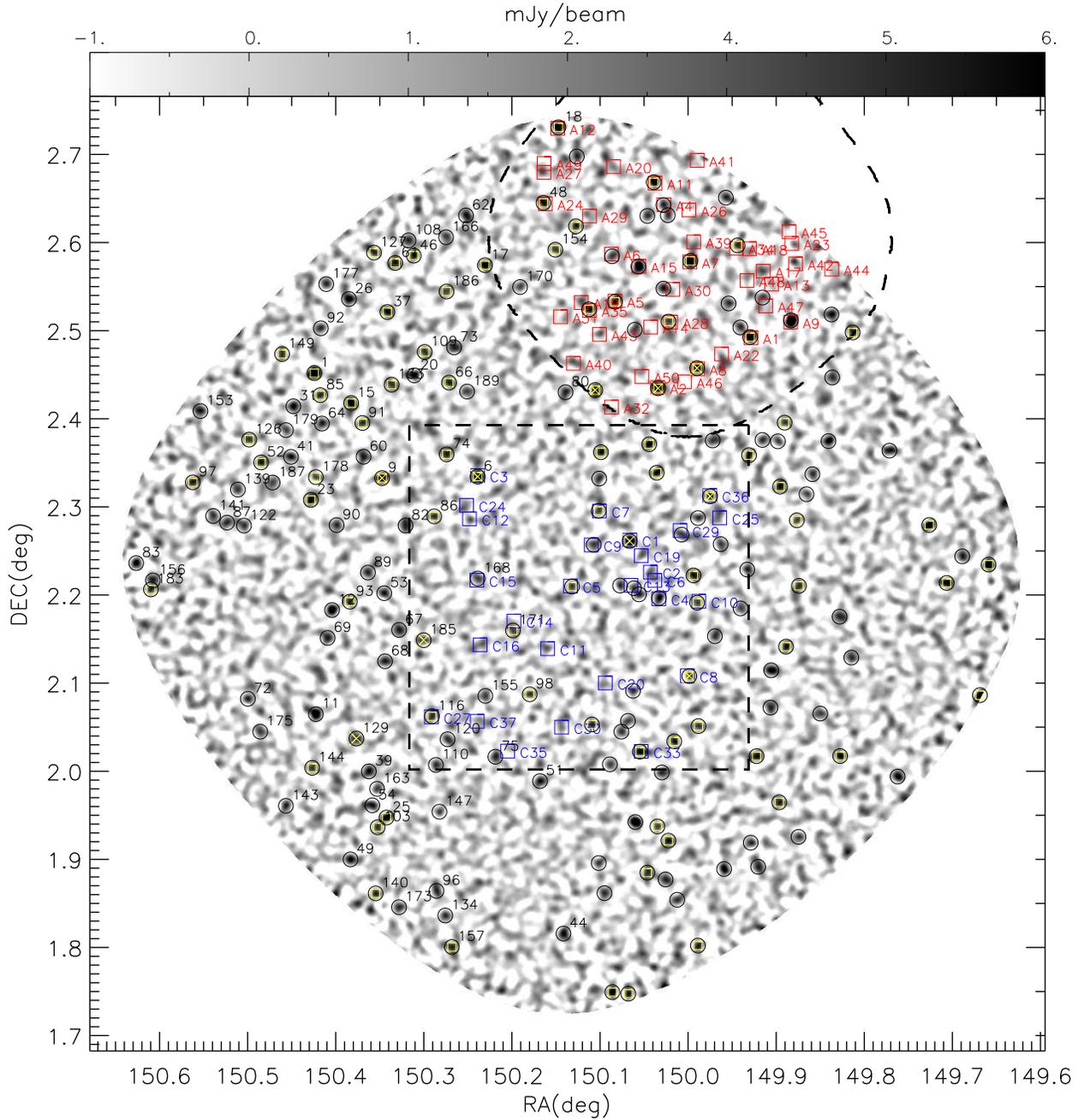,width=1.1\hsize,angle=90}
\caption{
AzTEC map, and comparison with other mm catalogs. Circles and numbers
denote AzTEC sources from this work.  Blue and red squares denote
COSBO (Bertoldi et al. 2007) and AzTEC/JCMT (Scott et al. 2008)
sources, with accompanying labels C\# and A\# that refer to their
catalog numbers, respectively. The boundaries of the surveys are
marked with dashed lines. The AzTEC sources from this work also
detected by the Bolocam survey are marked with crosses (Aguirre et
al., private communication). AzTEC sources from this work robustly
associated with 1.4GHz sources in this or other works are marked with
inner yellow squares. }
\label{fig:map0805}
\end{figure*}

The AzTEC/ASTE map includes areas of the sky previously imaged by other
mm-wavelength surveys, including COSBO (Bertoldi et al. 2007) and
AzTEC/JCMT (Scott et al. 2008). Figure~\ref{fig:map0805}  represents 
the catalogs of these two surveys overlaid on
the AzTEC/ASTE map, in order to emphasize common sources and differences
among catalogs. Also shown are the unpublished sources detected
by Bolocam at the CSO (J. Aguirre, private communication) that
are common to the AzTEC catalog.

The AzTEC/ASTE and AzTEC/JCMT surveys have similar noise properties with
rms~$\sim1.25$~mJy~beam$^{-1}$. The 10m ASTE image, however,  has a
resolution of 34~arcsec, while the JCMT image has an effective resolution of
18~arcsec, after Wiener filtering. Due to the considerable level of
incompleteness at low-S/N in both catalogs, not all sources
are expected to be found in both. Out of the 50 candidate
sources extracted from the AzTEC/JCMT survey at S/N$\geq 3.5$,
48 fall within an
overlapping region with the AzTEC/ASTE survey. In
order to find which entries in the catalogs match each other, we adopt
the positional uncertainty relation that depends on the effective beam
size and S/N of the detections (\S~\ref{sec:posuncertainty}). This relationship
has been shown to work well for both the AzTEC/JCMT (Scott et al. 2008) and
ASTE data (Fig.~\ref{fig:astrom}). Since a real source will
suffer from positional uncertainties in both catalogs, we will use for
each ASTE source a search radius at the 95~per cent confidence level
of containing the location of the real source, and
will add in quadrature the 95~per cent confidence positional
uncertainty radius of a potentially matching JCMT source.
If the
distance between the JCMT and ASTE catalog positions are smaller than the
resulting search radius, both catalog entries will be considered to
correspond to the same source.
Search radii derived in this manner range from 6.4 to 13.6~arcsec.
A candidate source that is found
in two independent datasets also increases its individual reliability
over the average false-detection rate corresponding to its nominal S/N.
Out of the 48 AzTEC/JCMT source candidates that fall within the ASTE surveyed
area, 16 are common to the ASTE catalog, and are listed in
Table~\ref{tab:stacks}.

The coincidence between the bright AzTEC/ASTE and AzTEC/JCMT sources
is remarkable. From the list of the 15 brightest AzTEC/JCMT sources
that were followed up and confirmed with SMA (Younger et al. 2007,
2009), only AzTEC 13 and 14 are not found in the AzTEC/ASTE
catalog. Both sources have deboosted flux densities $S_{1.1\rm mm}
\approx 4.4\pm 1.3$~mJy (Scott et al. 2008). At these flux densities,
the completeness estimated for the AzTEC/ASTE map indicates a $\sim
57$~per cent chance of recovery. The other two sources in this flux
range, AzTEC/JCMT 11 and 12, are indeed detected. The level of
recovery at lower flux densities is lower, as expected by the
completeness function. 

\begin{figure}
\epsfig{file=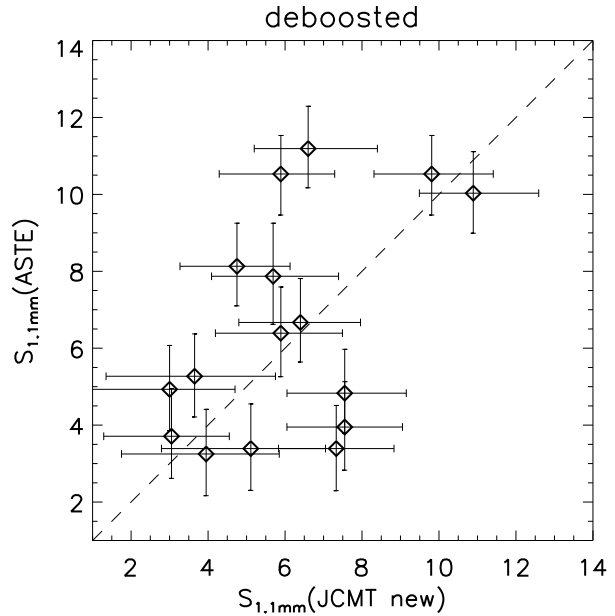,width=1.05\hsize,angle=90}
\caption{ Deboosted ASTE versus deboosted JCMT fluxes corresponding to the
common sources between this paper and the revised catalog published in
Downes et al. (2011). Error bars represent 68~per cent confidence intervals 
in deboosted fluxes.
}
\label{fig:ASTEvsJCMT}
\end{figure}

 Figure~\ref{fig:ASTEvsJCMT} represents the
deboosted fluxes of the AzTEC/ASTE map versus the deboosted fluxes of
the AzTEC/JCMT revised catalog (Downes et
al. 2011). The filtering and deboosting methods employed were exactly
the same for both datasets. The mean ratio between
fluxes is 1.08, with an r.m.s. of 0.44. Folded into these dicrepancies
are both the increased noise in the new filtering technique, blending 
of fainter sources within the larger ASTE beam and the impact of 
these effects on the deboosting process.

At 1.2mm the COSBO survey derived a catalog of 27 candidate sources
detected at $S_{1.2\rm mm} \gsim 2.2$~mJy, out of which 12 have
AzTEC/ASTE 1.1mm counterparts, following a positional uncertainty
analysis similar to that performed above. The FWHM for the MAMBO map
has been adopted to be 11~arcsec, and the resulting search radii at
95~per cent confidence level range between 6.2 and 14.0~arcsec. Common
sources are listed in Table~\ref{tab:stacks}.  The mean deboosted flux
density ratio for these common sources is $S_{1.1 \rm mm} / S_{1.2 \rm
mm} = 0.93\pm 0.43$.  Among the 15 brightest COSBO sources, with
$S_{1.2\rm mm} \gsim 4.4$~mJy, COSBO ID numbers 2, 6, 11, 12 and 14
are not formally detected in our AzTEC/ASTE catalog, although they
show $2.4\lsim {\rm S/N} \lsim 3.4$ peaks in the AzTEC S/N map. For instance,
COSBO source 2, $S_{1.2\rm mm} \gsim 5.9$~mJy (Bertoldi et al. 2007),
has a probability of detection in the AzTEC/ASTE map of $\sim 90$~per cent
(assuming a 0.93 mean flux ratio), and, indeed, has a S/N$\approx 3.0$
in the AzTEC/ASTE map at the COSBO position.

From the catalog of 19 candidate sources extracted by the Bolocam team,
10 are also detected by the deeper
AzTEC/ASTE survey, and are identified in Table~\ref{tab:stacks}. The
adopted effective FWHM for the Bolocam map in the positional
uncertainty calculation to match AzTEC/ASTE sources is 31~arcsec
(Laurent et al. 2005),
and the resulting search radii range between 10.9 and 18.7~arcsec.

\subsection{Overlap with the deep COSMOS radio-survey}

The VLA 1.4GHz deep mosaic of COSMOS (Schinnerer et al. 2010) also
provides an excellent catalog to look for counterparts of the SMGs
presented in this paper.
The tight correlation between radio continuum
emission, which is dominated by synchrotron radiation from supernova
remnants, and FIR emission dominated by thermal radiation from warm
dust heated by young stars, in galaxies (Helou, Soifer \&
Rowan-Robinson 1985, Condon 1992, Yun, Reddy \& Condon 2001)
translates into a large percentage of cross identifications among
catalogs derived in both frequencies (e.g. Ivison et al. 2002,
2007).

The matching process was carried out within 17~arcsec radii circles of
the AzTEC positions, looking for possible associations. To quantify
the significance of the possible associations, we have used the
$P$-statistic (Downes et al. 1986) which calculates the probability
that a radio source of a given observed flux density could lie at the
observed distance from the AzTEC source by chance. Only possible radio
counterparts with values of $P\leq 0.05$ are considered robust, and
are listed in Table~\ref{tab:stacks}, together with their distance
offsets and $P$-values.

Out of the 189 AzTEC source candidates, 77 (40~per cent) have
robustly associated radio counterparts within the COSMOS radio
catalog, and out of these 77, 7 (10~per cent) have a double robust
association. These percentages are lower than those of other SMG
studies ( $\sim 60-70$~per cent, e.g. Ivison et al. 2007, Biggs et
al. 2011) due to the large incompleteness at the 1.4GHz catalog limit
(a $4\sigma$ cut threshold is used) and shallower
nature of the 1.4GHz COSMOS data.
This is also the
reason why some previously claimed associations (see Table~1) are not
identified in this study, as they were based on lower threshold
detections.

\subsection{Multiwavelength photometry}

Table~\ref{tab:stacks} summarizes the photometry derived from the
comparison with other sub-mm to radio-wavelength catalogs.   The
1.4GHz flux densities of the robustly associated counterparts are one
to two order of magnitude lower than the deboosted 1.1mm flux
densities, with the exception of a few marked extended sources (see
below), suggesting the mm flux has a thermal origin rather than the
hyper-luminous synchrotron dominated blank-field sources identified by
the South Pole Telescope (Vieira et al 2010) or the
variable $S_{\rm 1.2mm}>10$~mJy flat-spectrum quasars discovered in
MAMBO fields (Voss et al. 2006). None of the bright 
$S_{\rm 1.1mm} {\rm (db)} \gsim
8$~mJy AzTEC sources are associated with luminous X-ray objects
either (Johnson et al. in prep), suggesting that, at best, they might 
harvour weak active galactic nuclei or be Compton-thick.

 In thermal dominated sources, the millimeter to radio flux
density ratio can be exploited as a redshift indicator. It increases
monotonically with redshift, with some degeneracy due to the variety
of radio synchrotron-slopes and mm dust-emissivity indices present in
the interstellar medium of those local galaxies used to define the
relationship (Carilli \& Yun 1999, 2000).  Additionally there exists a
level of degeneracy between the temperature of the dust generating the
rest-frame FIR luminosity (and hence mm flux) and the
redshift. Regardless, by adopting a library of local galaxy templates,
and accepting the intrinsic dispersion in their SEDs, the 1.4GHz to
1.1mm flux-density ratio still provides a crude but useful estimation
of the redshift (Carilli \& Yun 2000, Rengarajan \& Takeuchi 2001,
Aretxaga et al. 2007).  This indicator becomes relatively insensitive
to redshifts beyond $z\sim 3$, as the 1.1mm filter starts to sample
the flattening of the SED towards the rest-frame FIR peak, whilst
still providing a powerful discriminant between low-redshift ($z < 2$)
and high-redshift ($z > 2$) objects.

\begin{figure}
\epsfig{file=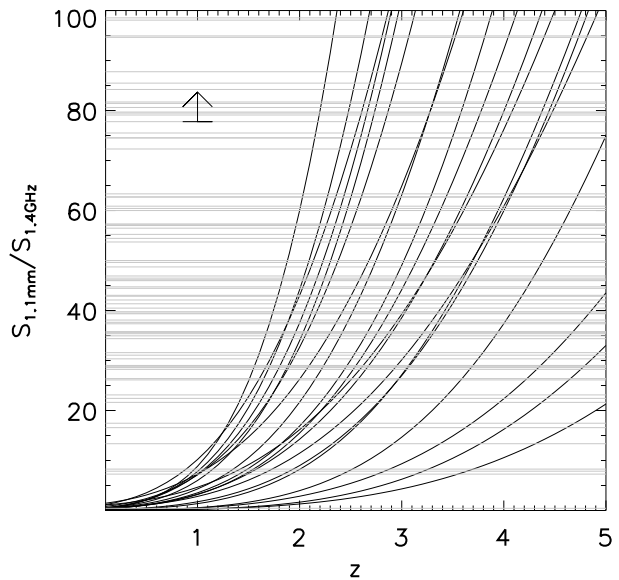,width=1.05\hsize,angle=90}
\caption{Millimeter to radio flux density ratio as a function of
redshift for 20 local galaxies used as templates in the derivations
of photometric redshifts of SMGs (Aretxaga et al. 2007). The grey horizontal
lines represent the colours of the AzTEC galaxies with robustly associated
radio counterparts, as in Table~1. The arrow at the top left corner
represents the typical $3\sigma$ lower limit for SMGs in the COSMOS map
with undetected radio counterparts.
}
\label{fig:photom}
\end{figure}

Figure~\ref{fig:photom} represents the colour-redshift diagram of 20
local galaxies used as templates in the derivations of photometric
redshifts (Aretxaga et al. 2007) and the colours measured for the
AzTEC sources with radio associations. Except for a few exceptions
(like AzTEC/C61, a likely radio-loud AGN, or C124 which is associated
with a likely foreground radio galaxy at $z\sim 0.3$) that appear at
the bottom of the diagram, the majority of the SMGs have mm-to-radio
colours indicative of $z>1$ systems, in accordance to the expectations
derived from other (sub-)mm-wavelength redshift distributions for the
population (Chapman et al. 2003, 2005, Aretxaga et al. 2003, 2007,
Pope et al. 2005, Valiante et al. 2007). A detail analysis of radio
and optical-IR counterpart associations to these AzTEC sources 
and their inferred redshifts is deferred to an upcoming publication.

\section{Number counts}
\label{sec:num_counts}

We derive estimates for the number density of SMGs as a function of
flux density, the so-called ``number counts'' using the Bayesian
technique originally outlined in Coppin et al. (2005, 2006) and used
extensively in previous AzTEC publications (for example, see
Austermann et al. 2009). 
 While other techniques commonly used for the extraction of source
counts from submm/mm wavelength surveys, in particular the "$P(D)$"
approach (e.g. Patanchon et al. 2009, Glenn et al. 2010), can in
principle estimate the counts at fainter flux densities (i.e. below
the detection limit of the survey), these methods are highly dependent
on the assumed model, and the formally derived errors do not always
represent the true uncertainty in the counts at faint flux densities
(e.g. see discussions in Scott et al. 2010 and Glenn et al. 2010). On
the other hand, with the Bayesian technique, the estimated counts are
only weakly dependent on the assumed model of the prior distribution
(Austermann et al. 2009, 2010), and the derived error bars more
accurately describe the uncertainty in this estimate. For this reason,
we use the Bayesian method, and derive the source counts only
down to a flux density limit of $S$(1.1mm) = 1.5 mJy.

We parameterize the number counts using a
Schechter function:
\begin{equation}
  \frac{dN}{dS} = N_{3\rm{mJy}}\left(\frac{S}{3\rm{mJy}}\right)^{\alpha+1}
                  e^{-(S-3\rm{mJy})/S^{\prime}}
\label{eq:schechter}
\end{equation}
with $N$ being the number of sources
per square degree, $S$ being the source flux density at 1.1mm wavelength, and
$\alpha$ being the power law slope of the faint-end counts.  In this
formalism $N_{\mathrm{3mJy}}$ has a natural interpretation as
the number of sources per square degree with a flux of 3~mJy.

The measured flux densities of sources blindly detected in the AzTEC
map must be corrected for biases resulting from the interaction
between the Gaussian noise distribution in the map and the underlying
flux density distribution of sources on the sky.  We perform this
correction by constructing the full posterior flux density
distribution for each source, taking as a prior the parameters
$N_{3\rm{mJy}}=160$ sq. deg.$^{-1}$, $S^{\prime}=1.3$~mJy and
$\alpha=-2.0$, which are consistent with those measured in the SHADES
fields (Austermann et al. 2010). We nevertheless iterate on the
adopted prior values to guarantee weak dependence on the starting
values for the end results.
For the $\geq 3.5\sigma$ peaks in the map, the full posterior
probabilities are parameterized by their maxima and 68~per cent
confidence levels, and are listed as deboosted fluxes in Table~1.

Once the full posterior flux distribution is derived for all
$\geq 2.5\sigma$ peaks in the map, we cut all
source candidates whose posterior flux distribution indicates a 5~per cent or
greater probability of having a negative intrinsic flux.  This
is likely a very strict cut on our catalog,
however, the larger beam-size and greater confusion in this survey
warrants a conservative first approach (Austermann et al. 2009).

\begin{figure*}
\begin{tabular}{cc}
\epsfig{file=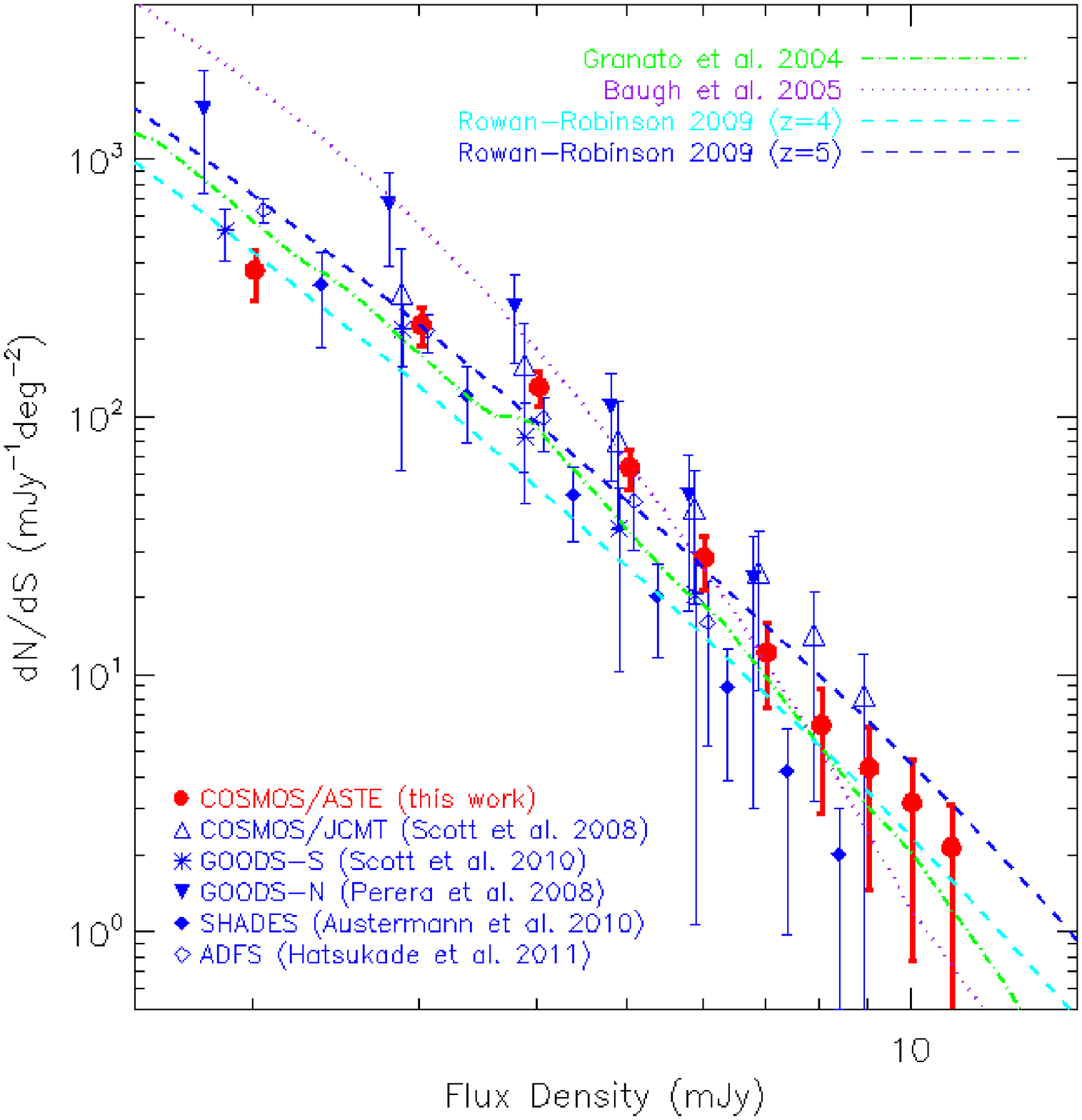,width=0.5\hsize}
\epsfig{file=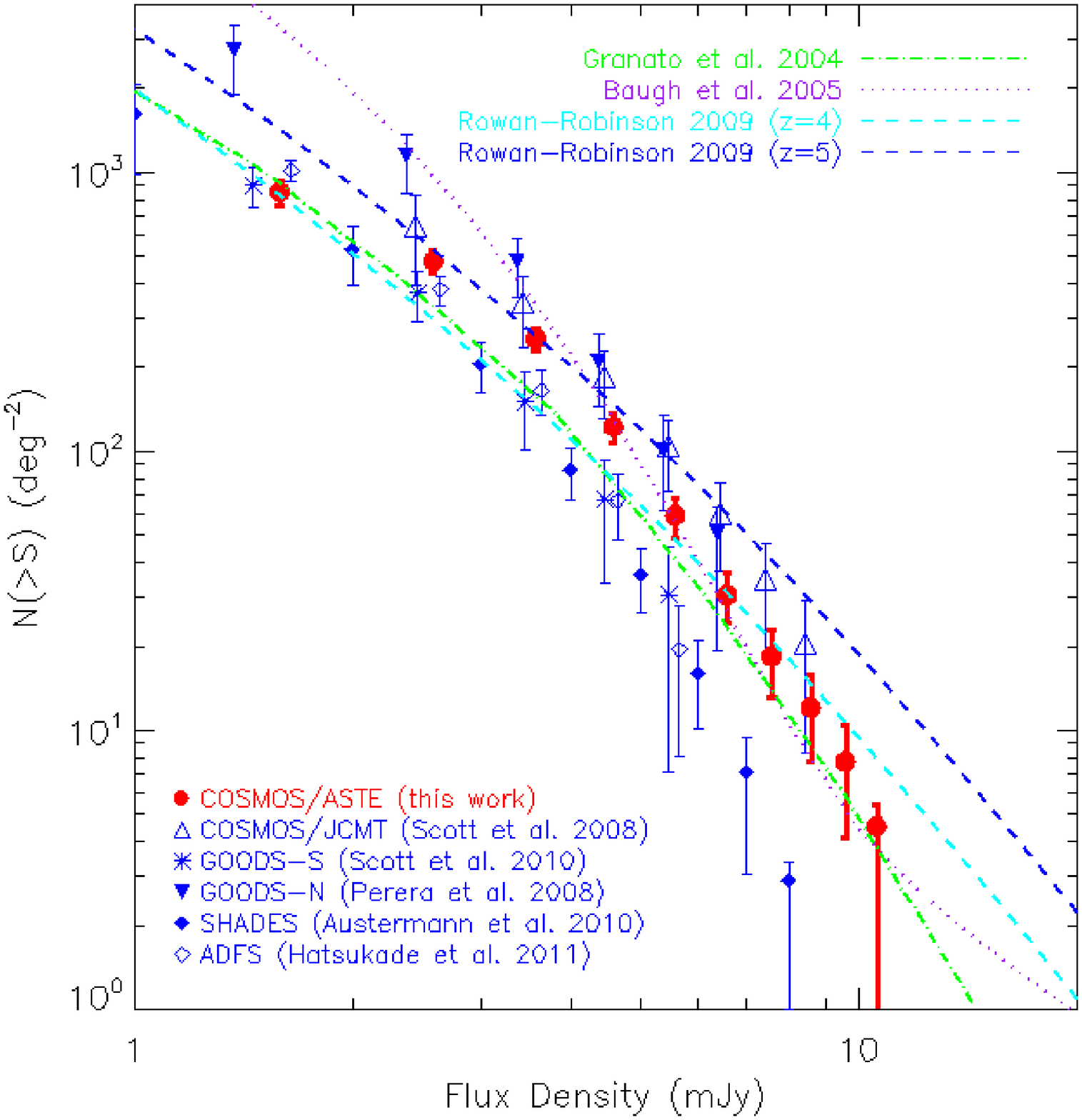,width=0.5\hsize}
\end{tabular}
\caption{(Left:) Differential number counts for the COSMOS AzTEC/ASTE field
  (red) along with the re-analyzed counts from previously published
  AzTEC studies.  Because of its large field, the COSMOS AzTEC/ASTE
  counts are more sensitive than previous studies in the
  3-15~mJy range.
  Overplotted (lines) are a number of 1.1mm number count
  predictions from an array of semi-analytic models for galaxy formation,
where $z$ for the Rowan-Robinson (2009) models denotes the 
free parameter that
describes the onset of the IR-luminous phase.
  (Right:) Corresponding integrated number counts at 1.1mm.
}
\label{fig:numcounts_all}
\end{figure*}

We bin our resulting catalog in 1~mJy flux density bins, correct each
bin for the corresponding completeness, and calculate $dN/dS$
uncertainties by bootstrap sampling the $dN/dS$ probability
distribution in each bin 20,000 times, taking also into account the error
in completeness.  Table~2 lists the resulting bin centers,
differential number counts, and 68~per cent confidence interval
uncertainties and figure~\ref{fig:numcounts_all} shows the differential
and integrated number counts derived for the full 0.72~sq. deg. COSMOS
field in this manner.
Also plotted are the number counts for all other published AzTEC extragalactic blank-fields (Scott et al. 2008, Perera et al. 2008,
Austermann et al. 2010, Hatsukade et al. 2011)
following a re-analysis of each of these maps using the
same technique outlined in section~3, in order to ensure data processing
effects take no role in the differences found.

Overall, the counts from different fields show some striking variance,
especially if one focuses on the two largest, COSMOS and SHADES:
the 0.72~sq. deg. COSMOS field with systematically more sources at
all flux levels than the  0.5~sq. deg. (to approximately the same noise level) SHADES
fields. The result however, is not as extreme as that shown in the comparison
with the smaller 0.15~sq. deg. AzTEC/JCMT COSMOS field.
In sections~7 and 8 we will explore further these differences.

\begin{figure}
\epsfig{file=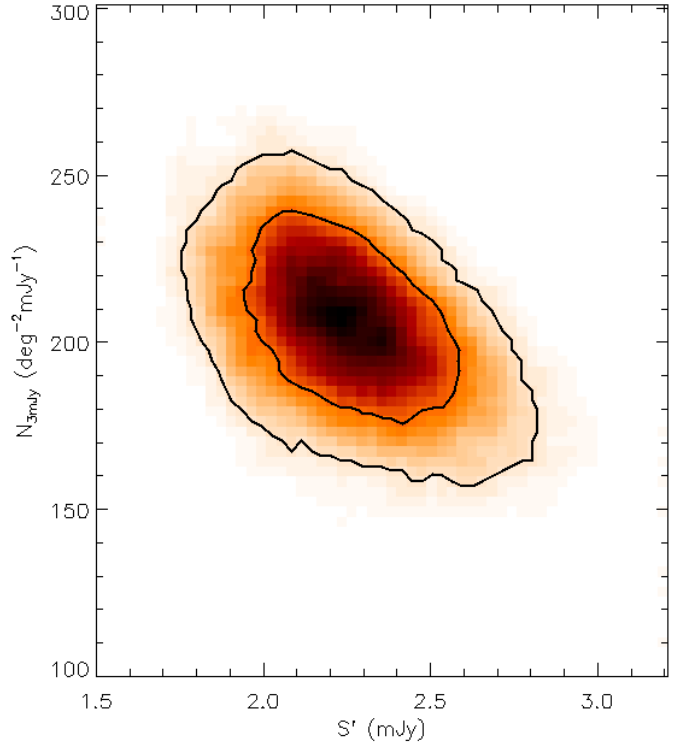,width=1.1\hsize}
\caption{Confidence limits for $N_{3\rm{mJy}}$ and $S^{\prime}$ when
  holding $\alpha=-2$.  Contours represent the 68~per cent and 95~per cent
  confidence limits.}
\label{fig:nc_fit_contours}
\end{figure}

We derive best-fit parameters for the Schechter function that
describes the COSMOS/ASTE differential number counts
by fitting Eq.~1 with a
Levenberg-Marquardt least-squares algorithm that uses the full data-data
covariance matrix in the $\chi^2$ calculation.  Our data does not
meaningfully constrain the faint end of the counts and thus we fix the
faint-end power-law index, $\alpha$, to a value of $-2$. This will also allow
a direct comparison to similar fits in the literature.
The best fit values of $N_{3\rm{mJy}}$ and $S^{\prime}$ for COSMOS are
given in Table~3 and
Figure~\ref{fig:nc_fit_contours} represents the error contours for
these parameters.

\begin{table}
\small
\label{table:nc_results}
\caption{COSMOS/ASTE differential and integral number counts, calculated as described in the text. The differential number counts
flux bins are 1-mJy wide with effective bin centers (first column)
weighted according to the assumed prior.}
\vspace{-4pt}
\begin{center}
\begin{tabular}{cccc}
\hline
$S$ & $dN/dS$             & $S$ & $N(>S)$ \\
mJy        & mJy$^{-1}$deg$^{-2}$ & mJy         & deg$^{-2}$ \\
\hline
 1.41 & $394^{+116}_{-140}$& 1.00 & $1038^{+132}_{-157}$\\
 2.44 & $269^{+54}_{-60}$& 2.00 & $644^{+63}_{-70}$\\
 3.44 & $176^{+28}_{-31}$& 3.00 & $375^{+34}_{-38}$\\
 4.45 & $99.5^{+15.0}_{-17.2}$& 4.00 & $199^{+19}_{-22}$\\
 5.45 & $49.9^{+8.9}_{-10.1}$& 5.00 & $99.1^{+11.3}_{-13.8}$\\
 6.46 & $22.3^{+5.0}_{-6.4}$& 6.00 & $49.1^{+7.0}_{-9.4}$\\
 7.46 & $10.3^{+3.4}_{-4.3}$& 7.00 & $26.9^{+4.8}_{-6.9}$\\
 8.46 & $5.83^{+2.33}_{-3.33}$& 8.00 & $16.63^{+3.47}_{-5.39}$\\
 9.46 & $4.07^{+1.80}_{-2.82}$& 9.00 & $10.79^{+2.57}_{-4.24}$\\
 10.46 & $2.94^{+1.40}_{-2.33}$& 10.00 & $6.72^{+1.83}_{-3.17}$\\
 11.46 & $1.87^{+0.87}_{-1.87}$& 11.00 & $3.78^{+1.18}_{-2.15}$\\
\hline
\end{tabular}
\end{center}
\end{table}

\begin{table}
\small
\label{table:nc_fit_params}
\caption{Best-fit Schechter function parameters to the COSMOS AzTEC/ASTE
  differential number counts with alpha fixed to $-2.$}
\vspace{-4pt}
\begin{center}
\begin{tabular}{ccc}
\hline
{\bf $N_{3\rm{mJy}}$}  & {\bf $S^{\prime}$} & {\bf $\alpha$} \\
deg$^{-2}$        &  mJy               & \\
\hline
$207^{+18}_{-20}$      &  $2.25 \pm 0.20$  & $-2$\\
\hline
\end{tabular}
\end{center}
\end{table}

\section{AzTEC sources vs.  Large Scale Structure in the COSMOS field}
\label{sec:LSScorr}

With the detection of ultra-bright SMGs by AzTEC (Wilson et al.
2008b, Ikarashi et al. 2011), MAMBO (Lestrade et al. 2010),
the South Pole Telescope
(Vieira et al. 2010) and the {\it Herschel} satellite (Negrello et al. 2010),
there has been new attention given to the role that lensing plays in
our view of the SMG population.  Our AzTEC survey of the COSMOS field
offers a prime opportunity to inspect the relationship between the
background SMGs  and the foreground large scale structure mapped out at
optical-IR wavelengths  over a large redshift span.
Indeed, with the 0.15~sq. deg. AzTEC/JCMT map of COSMOS Austermann et al.
(2009) already found a significant spatial correlation
between the projected foreground ($z\lsim1$) galaxy population and the
50 SMG candidates.
In this section we will take advantage of a much wider range of
foreground large scale structures covered within the
0.72~sq. deg. AzTEC/ASTE map of COSMOS
in order to test if the foreground structure
significantly impacts our view of the blank-field SMG population and 
whether this might be the likely origin of the difference 
in number counts between 
COSMOS and SHADES. 

\begin{figure}
\epsfig{file=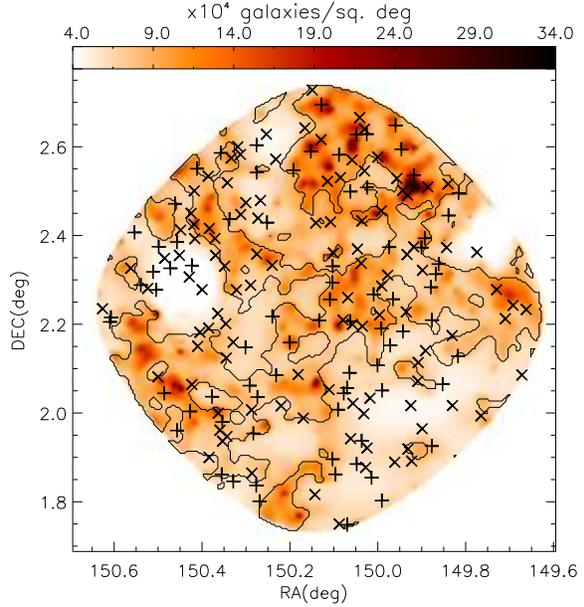,width=1.0\hsize,angle=90}
\caption{Smoothed surface-density map of galaxies with photometric
  redshifts $z<1.1$ derived from the
  optical-IR catalog of COSMOS (Scoville et al. 2007b), updated to
  include photometric redshifts derived from 30 optical-IR bands
  (Ilbert et al. 2009). Only the 0.72 sq. deg area surveyed by AzTEC
  with uniform ($\geq 50$~per cent) coverage is represented. Darker
  colours indicate more densely populated areas of the sky. The cross
  and plus symbols represent AzTEC sources detected at signal-to-noise
  ratios S/N$\ge4$ and $4>{\rm S/N}\ge3.5$, respectively. The contours
  divide the map into zones with lower or higher than the average
  density of optical-IR galaxies. The total areas of these zones are
  0.392 sq. deg. and 0.329 sq. deg, respectively. }
\label{fig:galden1}
\label{fig:twotone}
\end{figure}

Figure~\ref{fig:galden1} shows the projected density of optical-IR
galaxies in the AzTEC field with photometric redshift
$z_{\rm{phot}}\leq 1.1$ along with the location of the 129 AzTEC sources
with S/N $ \geq 4.0$, $\lsim 2$~per cent of which we expect to be
spurious detections.  While there is no apparent correlation between
the positions of both populations, we can quantify this impression
by comparing the distribution of projected densities of
optical-IR galaxies within 30~arcsec radii circles centered on the
AzTEC positions with those of 30~arcsec radii
circles centered on random locations in the map (as in Austermann et al. 2009).  A Kolmogorov-Smirnov
(KS) $D$-test can then be used to ask with what probability we can
rule out the null hypothesis that the two samples are drawn from the
same parent distribution.  As expected, the result, $P_{\rm
KS}=68$~per cent, is consistent with no strong correlation between the
optical-IR population at $z\leq 1.1$ and the $S/N\geq 4$ AzTEC catalog.

Performing the same test with a flux-cut on the AzTEC catalog results
in a more significant correlation detection.  Considering only the 41
S/N $\geq 5$ sources we find $P_{\rm KS}=1.8$~per cent, which is
tentative evidence (at $\approx 2.4\sigma$ level) that the null
hypothesis of no correlation is rejected.
Indeed, for the 20 AzTEC
sources with deboosted flux-densities $S_{\rm 1.1mm} \geq 6$~mJy we
can strongly reject the null hypothesis of identity between the
distributions of galaxy-densities around random positions in the AzTEC
coverage area of the COSMOS field and galaxy-densities around bright
AzTEC sources: $P_{\rm KS}=0.11$~per cent ($\approx 3.3\sigma$).

\begin{figure}
\hspace*{-0.7cm}
\epsfig{file=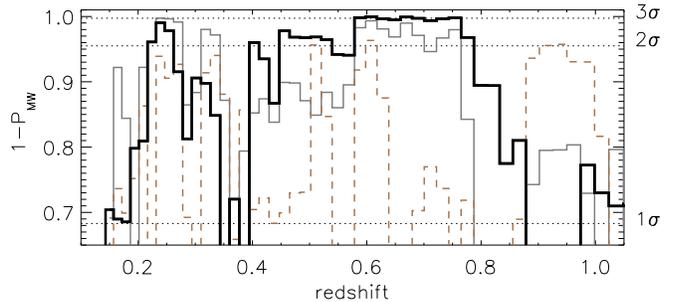,width=0.57\hsize,angle=90}
\caption{ Probability that the
  median galaxy density around AzTEC sources
  is significantly   larger
  than the density of galaxies around random positions in the
  AzTEC-covered map versus redshift. 
  The thick black solid line represents probabilities around 
deboosted $S_{\rm 1.1mm} \geq 6$~mJy sources, the grey solid  line represents
probabilities around 
deboosted $S_{\rm 1.1mm} \geq 5$~mJy sources, and the dashed brown line
represents probabilities around $S/N \geq 4$ sources. 
{The redshift bins have an increasing step ranging from $\Delta z =0.014$ 
to 0.026, at $z=0.15$ to 1.1, to accommodate a sufficient sample of galaxies and the increasing degradation in photometric redshift precission
(e.g. Scoville et al. 2007b)}
}
\label{fig:z_contrib}
\end{figure}

The latest release of photometric redshifts for optical-IR selected
galaxies in the COSMOS field has achieved accuracies $\Delta z / (1+z)
\approx 0.007$ to $0.012$ at $z\lsim1.25$ (Ilbert et al. 2009).
This statistical precision allows for the
identification of large-scale structure pertaining to different
redshift slices (see for example Scoville et al. 2007b). With this
information in hand, we can search in redshift space for the
structures that are more likely associated with the bright AzTEC
sources. We will again compare the distributions of galaxy-densities
around AzTEC sources and around random positions in the AzTEC covered
COSMOS area at the redshift of interest. As a reference, if we take
into account the AzTEC catalog of 129 S/N$\geq 4$ sources, there is no
significant signal of statistical differences between the
distributions at any redshift (see Fig.~\ref{fig:z_contrib}).

However, if we restrict the analysis to the 20 sources with deboosted flux densities
$S_{\rm 1.1mm} \geq 6$~mJy, we identify the redshift ranges
$0.58 \lsim z \lsim 0.76$ and $0.21 \lsim z \lsim 0.26$
as the ones in which most significant ($\sim 3\sigma$) differences are found
between the median density of galaxies around AzTEC sources and the median density of galaxies
around random positions (Fig.~\ref{fig:z_contrib}).
The same redshift bins are highlighted if the analysis focuses on the 42 sources with
deboosted flux densities $S_{\rm 1.1mm} \geq 5$~mJy, although the probability of
rejecting the null hypothesis for the former redshift interval is smaller than for the later.

\begin{figure}
\epsfig{file=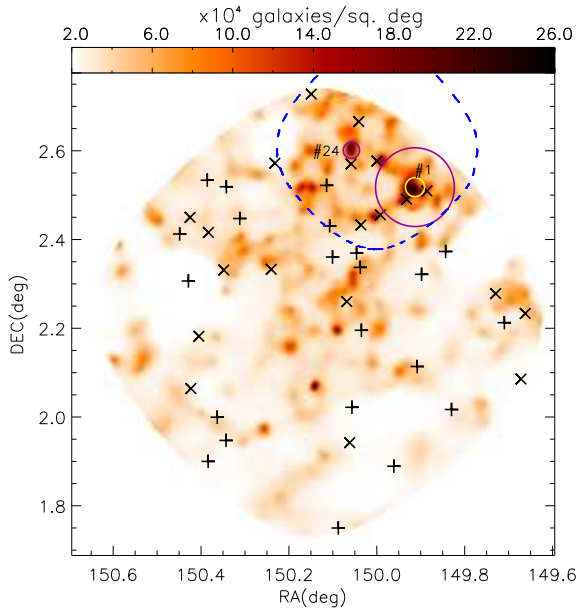,width=1.0\hsize,angle=90}
\caption{Smoothed surface-density map of galaxies at $0.58\lsim z \lsim0.76$
  detected at optical-IR wavelengths by the COSMOS survey within the uniform
coverage area of the AzTEC/ASTE map.
  The cross symbols represent 20 AzTEC
  sources detected with deboosted flux densities $S_{\rm 1.1mm} \geq 6$~mJy.   The plus symbols
denote sources detected with deboosted flux densities $5 \leq S_{\rm 1.1mm} < 6$~mJy.
  A massive cluster at $z\approx 0.73$ (Guzzo et al. 2007) located in the NW of the map is
marked as \#1. The inner yellow circle
of 1.5~arcmin diameter
 marks the core of X-ray emission, while the purple 6~arcmin diameter
circle marks the FWHM of the optical/IR overdensity associated with the cluster
(Scoville et al. 2007b). Another rich cluster at $z\approx 0.61$ from the large-scale
structure catalog of COSMOS (Scoville et al. 2007b) is marked as \#24, and the
FWHM of the optical/IR overdensity is encircled in purple.
The outer blue dashed-line contour depicts the edge of the
uniform coverage of the AzTEC map observed in 2005 at the JCMT
(Scott et al. 2008).
}
\label{fig:galden_slice0p65}
\end{figure}

\begin{figure}
\epsfig{file=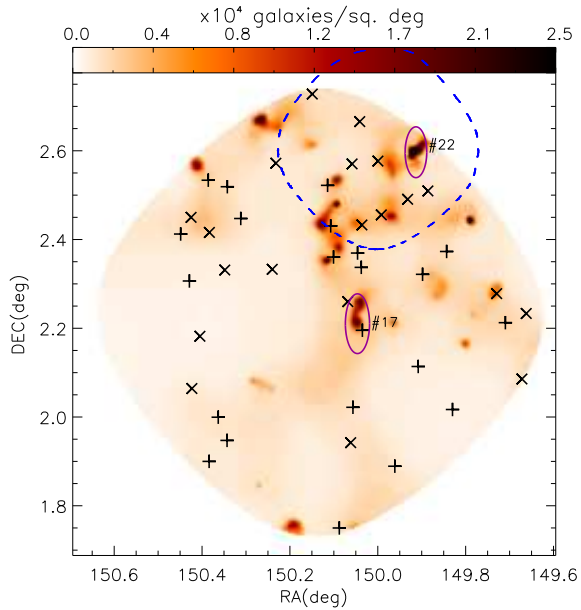,width=1.0\hsize,angle=90}
\caption{Smoothed surface-density map of galaxies at $0.21\lsim z \lsim0.26$
  detected at optical-IR wavelengths by the COSMOS survey within the uniform
coverage area of the AzTEC map observed in 2008.  Symbols and lines
are as in Fig.~\ref{fig:galden_slice0p65}.
Two X-ray clusters from the catalog of large-scale structures in COSMOS
(Scoville et al. 2007b) are marked as \#17 and \#22.
}
\label{fig:galden_slice0p24}
\end{figure}

The optical-IR galaxy density in these redshifts slices and the positioning
of the bright AzTEC sources can be seen in figures.~\ref{fig:galden_slice0p65} and \ref{fig:galden_slice0p24}. In the interval
$0.58 \lsim z \lsim 0.76$ we find  that $S_{\rm 1.1mm} \geq 6$~mJy AzTEC
sources
have a significantly denser galaxy environment than that found
at random positions in the map:
the null hypothesis of identity between the medians of the distributions can be rejected at a
$1-P_{\rm MW}=99.88$~per cent confidence level, while for $S_{\rm 1.1mm} \geq 5$~mJy AzTEC sources
that level gets reduced to $1-P_{\rm MW}=98.7$~per cent. In the interval
$0.21 \lsim z \lsim 0.26$ the reverse happens, finding that the
$S_{\rm 1.1mm}
\geq 6$~mJy AzTEC sources have a denser environment with a significance
$1-P_{\rm MW}=99.2$~per cent. The significance gets increased to
$1-P_{\rm MW}=99.95$~per cent
if $S_{\rm 1.1mm} \geq 5$~mJy sources are considered. The effect seems
to be carried by four $5 \leq S_{\rm 1.1mm} < 6$~mJy sources that coincide
with large density peaks, rather than by the general population of
intermediate/brightness sources, though.
If one excludes the $S_{\rm 1.1mm} \geq 6$~mJy
sources from the analysis, the probability drops to
$1-P_{\rm MW}=99.0$~per cent.

These two redshift intervals are the same ones we identified as
having the largest correlation between $z\lsim 1.1$ optical-IR
galaxies and AzTEC sources for the smaller 0.15~sq. deg. field
observed by AzTEC in the JCMT (Austermann et al. 2009). It is thus
important to know if the correlations arise mainly due to the very
rich clusters located in the area of the sky previously sampled, or if
this is a trend observed over the larger 0.72~sq. deg. field, which
might be more representative of a generic blank-field.

\begin{figure}
\hspace*{-0.8cm}
\epsfig{file=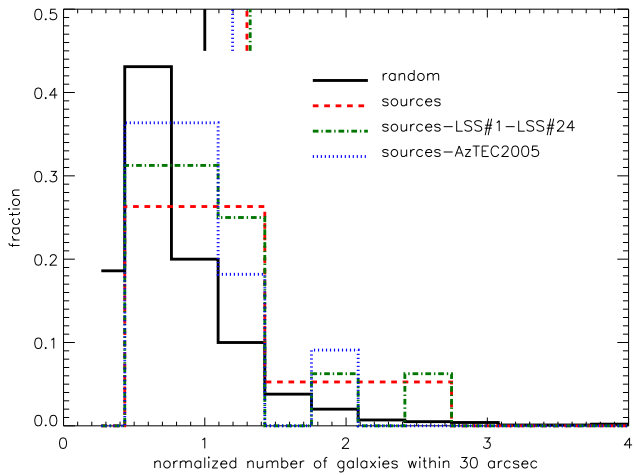,width=0.8\hsize,angle=90}
\caption{Histograms of the number of optical/IR galaxies
around random positions in the AzTEC covered COSMOS map, normalized
to the mean value over the whole map, represented a solid black line.
The dashed, dash-dot-dash and dash-dotted lines  represent the
distribution of number of galaxies found around AzTEC sources with deboosted
flux
densities $S_{\rm 1.1mm} \ge 6$~mJy in the entire AzTEC covered COSMOS map,
excluding the 6~arcmin diameter area around clusters \#1 and \#24,
and excluding
the whole uniform area mapped by AzTEC
in 2005 in the JCMT.
The vertical bars at the top of the diagram mark the mean values of the distributions.
}
\label{fig:dennocluster}
\end{figure}

In Austermann et al. (2009) we showed that the association of bright
AzTEC sources around areas of large galaxy densities was not
exclusively due to the presence of massive rich clusters. If we mask
out the areas marked as \#1 and \#24 in
figure~\ref{fig:galden_slice0p65}, which represent the FWHM of the two
massive clusters at
$0.58 \lsim z \lsim 0.76$ identified in the large-scale-structure
analysis of Scoville et al. (2007b), and exclude the sources that
fall within those areas, the probability of having a
median of galaxy-densities around AzTEC sources larger than that of
random positions by chance is only $P_{\rm MW} = 0.3$~per cent.
The distributions are shown in figure~\ref{fig:dennocluster}.
Likewise, if we exclude from the correlation analysis the
overlapping  area common to the AzTEC/JCMT and AzTEC/ASTE maps
(encircled in dashed blue lines in figures~\ref{fig:galden_slice0p65} and \ref{fig:galden_slice0p24}), the remaining 12 $S_{\rm 1.1mm} \geq 6$~mJy sources
still show a larger tendency of falling within the large galaxy-density
regions mapped at optical-IR wavelengths at $z\lsim 1.1$. In that case,
the null hypothesis of identity between the medians of the distributions
can be rejected at a $1-P_{\rm MW}=97.0$~per cent confidence level.

\section{Discussion}

\subsection{Effects of Foreground Structure on Number Counts}
\label{sec:fg_nc}

With evidence that our detected SMGs are spatially correlated with
foreground ($z\leq1.1$) large scale structure in the COSMOS field, we now
ask what effect, if any, this structure might have on our
estimation of the SMG number counts.  We address this by first
dividing the catalog into two sub-samples: SMGs that lie in regions of
the map with foreground galaxy densities larger than the average density
of the field and SMGs that lie in regions with foreground galaxy
densities smaller than the average galaxy density of the field.
Figure~\ref{fig:twotone} shows the two regions along
with the respective locations of the corresponding SMGs.

\begin{figure*}
\begin{tabular}{cc}
\epsfig{file=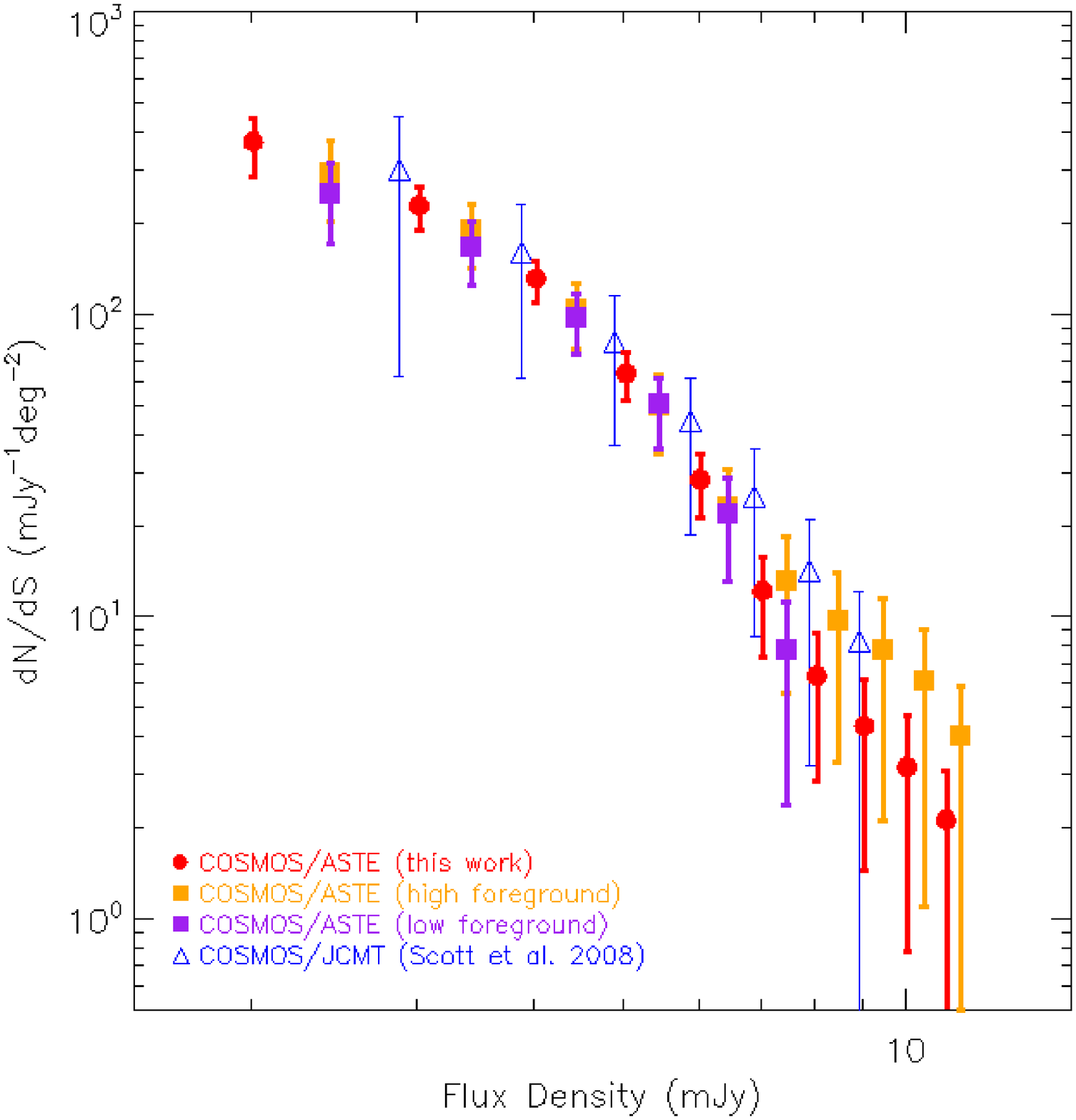,width=0.5\hsize}
\epsfig{file=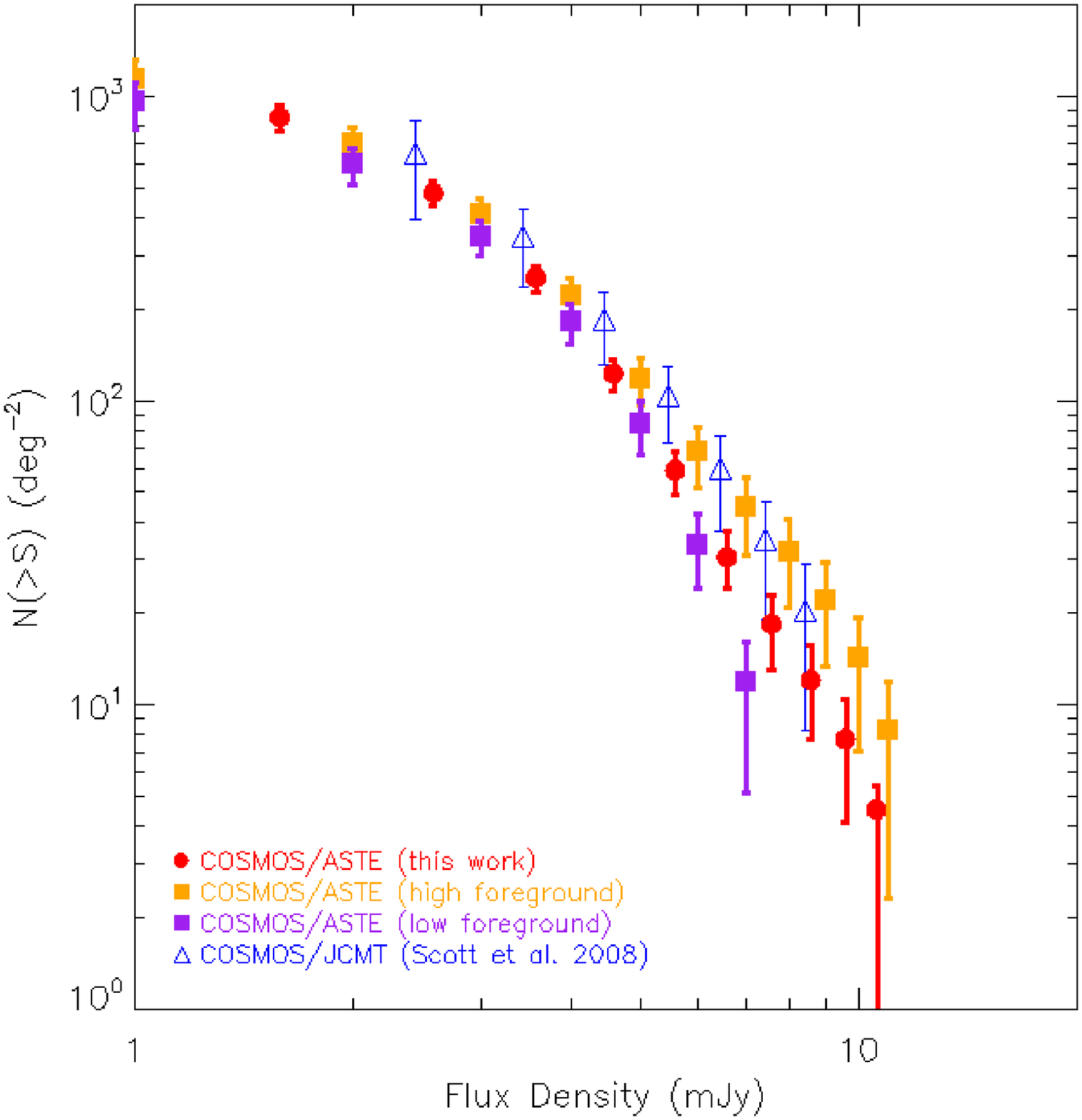,width=0.5\hsize}
\end{tabular}
\caption{(Left:) Differential number counts derived for regions of low and high galaxy density within the AzTEC/ASTE covered COSMOS area.   ``High
  foreground'' denotes sources that fall in regions of the map with
  greater than the mean foreground galaxy density.  ``Low foreground''
  denotes sources that fall in regions of the map with less than the
  mean foreground galaxy density. Triangles denote counts from the
  COSMOS/JCMT. (Right:) Corresponding integral number counts. }
\label{fig:compare_HL}
\end{figure*}

\begin{figure}
\hspace*{-0.8cm}
\epsfig{file=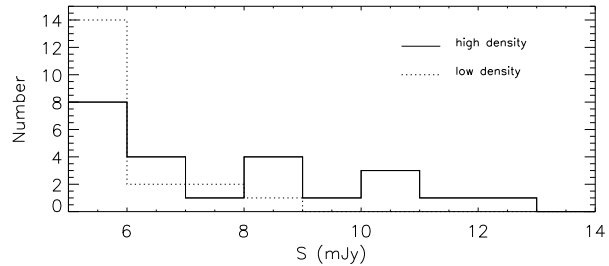,width=0.9\hsize,angle=90}
\caption{Distribution of fluxes for the 60 brightest ($S_{\rm 1.1mm} \geq
5$~mJy) AzTEC sources that fall within
high galaxy-density areas (continuous line) and within low galaxy-density
areas (dotted line) of the COSMOS map. The high density areas host the
majority of the brightest sources. The null hypothesis that the two distributions can be
derived from the same parent distribution can be rejected with a probability
$P_{KS}=0.37$~per cent.
}
\label{fig:dendencompa}
\end{figure}

Figure~\ref{fig:compare_HL} shows the number counts resulting
from splitting sources in high and low foreground galaxy density
environments.  While none
of the $dN/dS$ bins individually show a clear, statistically
significant deviation from the mean counts over the field, the data
clearly suggests that, consistent with the results of
Section~7, at high fluxes ($S_{\rm 1.1mm}\geq6 $mJy) the counts are
systematically higher for the high foreground galaxy density sample.
This result can also be made evident by representing the distribution of
flux densities of the SMGs that
fall within the low and high galaxy-density areas of the map
(see Fig.~\ref{fig:dendencompa}).
The null hypothesis that the 1.1mm flux-density distribution of sources that
fall within low and
high density areas are similar can be rejected, as differences as large
as the one measured can only be produced by chance in
0.37~per cent of situations.

The number counts for the low density galaxy sample is closer to the
number counts derived for SHADES, if still systematically above them (by
$\sim 16$~per cent in flux density), while the
high density galaxy sample
aligns with the results derived for COSMOS/JCMT and GOODS-N,
  (about
$\sim 30-40$~per cent offset in flux density at $S_{\rm 1.1mm} \gsim 5$mJy,
compare with Fig.~10). It is thus
apparent that an accurate
description of the overall population will fall somewhere in between
these solutions, and that despite having sampled 0.72~sq. deg. in
COSMOS and 0.5~sq. deg in SHADES at comparable noise levels, this is
not yet a large enough area to avoid variance due to intrinsic
clustering of the SMG population and amplification by foreground
structures.

 The optical-IR galaxies that compose
the COSMOS galaxy density map have accurate photometric redshifts that
place them at $z\le 1.1$, while most of the redshifts of the SMGs
uncovered by AzTEC are still unknown. Furthermore, the largest
amplitude correlation between the positions of optical-IR galaxies and
SMGs occurs at $0.6\lsim z\lsim 0.75$ (\S~7), where only a $\sim 3$~per
cent of 850$\mu$m SMGs with radio associations are statistically
located (Chapman et al. 2005).  Photometric arguments (Fig.~9,
\S~5.3), place the majority of AzTEC SMGs at $z\gsim1$ (see also
Younger et al. 2007, 2009) and the 4 bright targets with refined
interferometric positions by SMA that have been spectroscopically
targeted and have robust redshifts are indeed at
$z>1.1$ (Table~1, Capak et al. 2011 and in prep.,
Riechers et al. 2010, Smol\u{c}i\'c et al. 2011).
While there is no proof that all bright $S_{\rm
1.1mm}\geq 6$~mJy sources are at high redshifts, it is likely that
most of these sources are at $z>1$ and that their association with the
optical-IR galaxy large scale structure is through amplification, an
explanation already posed to account for the discrepancies between the
number counts measured in the smaller COSMOS AzTEC/JCMT field and
SHADES (Austermann et al. 2009).
Our new results, thus, confirm this
interpretation for a more representative area of the sky that is not
dominated by the presence of a rich cluster environment.

Lensing of galaxies by
foreground galaxies and by foreground groups of galaxies has been
shown to be the reason for the presence of a very bright (sub-)mm selected
galaxy population (Vieira et al. 2010, Negrello et al. 2010, Lestrade
et al. 2010, Ikarashi et al. 2011). These galaxies are extreme cases
of the phenomena presented here, where moderate $\sim 20$~per cent
amplification factors would be enough to account for the flux density
offset between the number counts of SHADES and COSMOS.

Light propagation experiments in cosmological simulations using
multiple lens-plane algorithms show that at $z>1$ any population is
subject to a large chance of amplification by foreground structures
(e.g. Martel \& Premadi 2008). Correlations between bright SMGs and
foreground optically selected galaxies at $z\sim 0.5$ were first
detected using a small sample of $S_{850\mu{\rm m}}\gsim 8$~mJy
sources in the UK 8-mJy and Hubble Deep Field (Almaini et
al. 2005).  The analysis of a statistically robust sample of 2477
350$\mu$m-selected SMGs in the Lockman-SWIRE field, that includes the
smaller UK 8-mJy survey area studied previously, has confirmed
correlations between the SMG population and  $z\sim 0.2$ 
and $z\sim 0.4$ optical and IR-selected
galaxy samples (Wang et al. 2011). Our first 0.15 sq. deg. survey
in the COSMOS field scanned all foreground structures at $0<z<1$,
yielding significant correlations between the bright 1.1mm-selected
SMGs detected in the field and optical-IR galaxies at redshifts $z\sim
0.25$ and $z\sim 0.65$ (Austermann et al. 2009). Whether these
correlations were dominated by the optical-IR overdensity of galaxies
where the AzTEC survey had been acquired or inherent of amplifications
to the general blank-field population was open to debate. Our new 
analysis in COSMOS allows for a better estimation of the structures
that contribute to the possible boosting of flux densities of SMGs by
foreground structures, identifying the $0.21\lsim z\lsim0.26$ and,
 most significantly, the $0.58\lsim z \lsim 0.76$ redshift bins
as those with the largest probability of association with bright
$S_{1.1{\rm mm}}\gsim 5$~mJy sources.  The first redshift interval
is common to the $350\mu$m-selected SMG correlations result. The
second redshift interval, however, is still unexplored by other
experiments.

Other populations of high-z galaxies, also should show similar
amplification trends. The positions of bright Lyman Break Galaxies
selected to be at $z\approx 2.5-5$ in the
Canada-France-Hawaii-Telescope Legacy Survey are, indeed, positively
correlated with optically-selected galaxies at $z<1.4$, and this
effect has been shown to be consistent with weak lensing by the
foreground structures in the line of sight (Hildebrandt, Waerbeke \&
Erben 2009). Strong lensing is also predicted to be a dominating
effect in the bright number counts of SMGs  ( Blain 1996,
Negrello et al. 2007, Lima, Jain \& Devlin 2010), however, neither
strong nor weak lensing are often included in the predictions of
observables offered by galaxy formation models.

\subsection{Comparison with galaxy formation models}

Figure~10 shows the number counts of COSMOS
compared with several semi-analytical galaxy formation models that have
successfully reproduced overall properties of the SMG and SMG
population (Granato et al. 2004, Baugh et al. 2005, Rowan-Robinson
2009). The 1.1mm number counts of the SHADES field were compared with
these very same models (Austermann et al. 2010), and it was
found that they all overpredicted the number counts
in the $\gsim 3$~mJy regime. The
discrepancies with the new 0.72~sq. deg. AzTEC/ASTE COSMOS field are
not that severe. While the Granato et al. (2004) model for the joint
formation of QSOs and SMGs could be made compatible with the COSMOS
number counts using a small shift in flux density that would mimic
the amplification claimed in this paper, the model over-predicts the
$S_{\rm 1.1mm}\lsim 2$~mJy number counts, which,
although they are poorly constrained by the COSMOS
data alone, have more robust estimations
from smaller deeper fields (Scott et al. 2010, Hatsukade et al. 2011).
The Baugh et al. (2005) model offers a good description of the COSMOS number
counts at $S_{\rm 1.1mm}\gsim 4$~mJy, but it over-predicts the number
counts at lower flux densities, while the Rowan-Robinson
(2009) models have a different functional form to that found for
COSMOS over the whole flux density range sampled by our study.

\section{Conclusions}

The number counts of the COSMOS 0.72~sq. deg. AzTEC/ASTE field show
an overdensity of sources with respect to the population of SMGs found
in previous large blank-field surveys such as SHADES (Austermann et
al, 2010). The number counts derived for the COSMOS field display a
systematic positive offset over those of SHADES, but are nevertheless
consistent with those derived from smaller fields that were considered
too small to characterize the overall blank-field population. We
identify departures to occur more significantly in the $S_{1.1 \rm mm}
\gsim 5$~mJy regime, and demonstrate that these differences are
related to the areas where galaxies at redshifts $z\lsim 1.1$ are more
densely clustered. The positions of optical-IR galaxies in the
redshift interval $0.60 \lsim z \lsim 0.75$ are the most strongly
correlated with the positions of the 1.1mm bright population ($S_{1.1
\rm mm}\geq 6$~mJy), a result which does not depend exclusively on
the presence of rich clusters within the survey sampled area.
The
most likely cause of these departures in number counts at 1.1mm is
lensing by either foreground galaxies or foreground groups of galaxies
at moderate amplification levels, that
increases in amplitude as one samples larger and larger flux
densities. Our results and the comparison with the previously
published SHADES number counts illustrate the fact that even $\sim
0.70$~sq. deg. surveys are still subject to variance due to the small
volume sampled by the mapped areas in conjunction to the chance amplification by foreground structures.

\section*{Acknowledgments}
This work has been supported in part by Conacyt (Mexico) grants
39953-F and 39548-F, NSF (USA) grants AST-0907952 and AST-0838222,
and by the MEXT Grant-in-Aid for Specially Promoted Research (No.~20001003).
Observations with ASTE carried out remotely from
Japan used NNT's GEMnet2 and its partner R\&E networks, which are
based on the AccessNova collaboration of the University of Chile, NTT
Laboratories, and the NOAJ.
We would like to thank everyone who supported
the AzTEC/ASTE observations of the COSMOS field, including
E. Akiyama, R. Cybulski,  K. Fukue, S. Harasawa, S. Ikarashi, H. Inoue,
M. Kawamura, A. Kuboi, J. Rand,  M. Tashiro, T. Tosaki, T. Tsukagoshi,
Y. Shimajiri and
 C. Williams. The ASTE project
is driven by the Nobeyama Radio Observatory (NRO), a branch of the
National Astronomical Observatory of Japan (NAOJ), in collaboration
with the University of Chile and Japanese institutions including the
University of Tokyo, Nagoya University, Osaka Prefecture University,
Ibaraki University, and Hokkaido University.

\end{document}